\documentclass[acmsmall]{acmart}
\usepackage{multirow}
\usepackage{longtable}
\usepackage{booktabs}
\usepackage{tabularx}
\usepackage{ragged2e}
\usepackage{tikz}
\usepackage{graphicx}
\usepackage{array}
\usepackage{caption} 
\usepackage{lipsum} 
\usepackage{makecell}

\renewcommand\footnotetextcopyrightpermission[1]{}

\AtBeginDocument{%
  }

\setcopyright{acmlicensed}
\copyrightyear{2018}
\acmYear{2018}
\acmDOI{XXXXXXX.XXXXXXX}

\acmJournal{JACM}
\acmVolume{37}
\acmNumber{4}
\acmArticle{111}
\acmMonth{8}

\begin{document}

%\title{A Review of Privacy-Enhancing Encryption Technologies in Data Sharing: With a Focus on Balancing Security, Performance, and Functionality}
\title{Privacy-Enhancing Encryption in Data Sharing: A Survey on Security, Performance and Functionality}
%\author{Xiaohong Li}
% \email{xiaohongli@tju.edu.cn}
% \affiliation{%
\author{Yongyang Lv}
\affiliation{
  \institution{Tianjin University}
  \city{Tianjin}
  \country{China}}
\email{lvyongyang@tju.edu.cn}

 \author{Xiaohong Li}
 \affiliation{
   \institution{Tianjin University}
   \city{Tianjin}
   \country{China}}
 \email{xiaohongli@tju.edu.cn}

 \author{Ruitao Feng}
 \affiliation{%
   \institution{Southern Cross University}
   \city{Gold Coast}
   \country{Australia}}
\email{ruitao.feng@scu.edu.au}
\authornote{Corresponding author.}

 \author{Xinyu Li}
 \affiliation{%
   \institution{Zhongguancun Laboratory}
   \city{Beijing}
       \country{China}}
\email{xinyuli1920@gmail.com}

\author{Guangdong Bai}
\affiliation{%
  \institution{City University of Hong Kong}
  \city{Hong Kong}
  \country{China}}
\email{guangbai@cityu.edu.hk}
\authornotemark[1]

\author{Leo Zhang}
 \affiliation{%
   \institution{Griffith University}
   \city{Queensland}
       \country{Australia}}
\email{leo.zhang@griffith.edu.au}

 \author{Lili Quan }
 \affiliation{%
   \institution{Singapore Management University}
   \city{Singapore}
       \country{Singapore}}
\email{liliquan@smu.edu.sg}

 \author{Willy Susilo}
 \affiliation{%
   \institution{University of Wollongong}
   \city{New South Wales}
       \country{Australia}}
\email{willy_susilo@uow.edu.au}

% \affiliation{%

% \affiliation{%

%\affiliation{%
%  \institution{Palmer Research Laboratories}


%% paper.
\begin{abstract}
The vigorous development of the Internet has spurred exponential data growth, yet data is predominantly stored in isolated user entities, hampering its full value realization. In large-scale deployment of ``AI+industries'' such as smart medical care, intelligent transportation and smart homes, the gap between data supply and demand continues to widen, and establishing an effective data sharing mechanism is the core of promoting high-quality industrial development. However, data sharing faces significant challenges in security, performance, and functional adaptability. Privacy-enhancing encryption technologies, including Attribute-Based Encryption (ABE), Proxy Re-encryption (PRE), and Searchable Encryption (SE), offer promising solutions with distinct advantages in enhancing security, improving flexibility, and enabling efficient sharing. Statistical analysis of relevant literature from 2020 to 2025 reveals a rising research trend in ABE, PRE and SE, focusing on their data sharing applications. Firstly, this work proposes a data sharing process framework and identifies 20 potential attacks across its stages. Secondly, this work integrates ABE, SE, PRE with 12 enhancement technologies and examines their multi-dimensional impacts on the security, performance, and functional adaptability of data sharing schemes. Lastly, this work outlines key application scenarios, challenges, and future research directions, providing valuable insights for advancing data sharing mechanisms based on privacy-enhancing encryption technologies.

\end{abstract}


%\ccsdesc[300]{Do Not Use This Code~Generate the Correct Terms for Your Paper}
%\ccsdesc{Do Not Use This Code~Generate the Correct Terms for Your Paper}
%\ccsdesc[100]{Do Not Use This Code~Generate the Correct Terms for Your Paper}



\maketitle

\section{Introduction}

In the contemporary era marked by the vigorous evolution of the Internet, data volumes are expanding at an exponential rate. A wide range of fields and scenarios, including the training of large-scale models, the advancement of scientific experiments, and the precise forecasting of market trends, are heavily reliant on substantial data resources \cite{[247],[54]}. Data, as a crucial factor of production, has been recognized as a key strategic asset by numerous countries worldwide \cite{[287],[288],[289]}. However, massive volumes of data remain siloed in private storage environments, which significantly impedes the full realization of data’s potential value \cite{[252]}. Only by establishing an effective data sharing mechanism to break down data silos can we unleash the latent value of data and drive innovation and development across various industries \cite{[249]}.\par
%Data sharing, a process where data owners provide their data resources to other users or systems, is extensively applied in fields like medical science \cite{[251],[252],[253]}, agriculture \cite{[256]}, social networks [248], and government affairs \cite{[249],[250]}.\par
%\feng{need to position this survey first if there is any relevant survey papers. Show what is expected according to the reality but not presented in their papers, emphersize the significance of the missing study dimensions (i.e., focus of our study). I found duplicated references. How are your works highlighted? Normally, people publish a survey to "advertise" the work from their group.}
%,[260]
%\newpage

%\feng{better positioning with other surveys right after the brief intro on the scope.}
Currently, the data sharing field is confronted with multiple complex challenges, which have significantly undermined data owners' willingness to engage in data sharing \cite{[310]}. Among these, the balance between security \cite{[245],[244]}, performance \cite{[32],[70]}, and functionality \cite{[101]} in the data sharing process is particularly prominent. Traditional encryption technology, while effective in ensuring data sharing security, has increasingly shown certain limitations as data sharing security requirements grow more complex \cite{[286]}. For instance, in multi-user scenarios where multiple data users require access to the same data, traditional encryption methods often depend on inefficient one-to-one encryption schemes or risk private key disclosure. In this context, privacy-enhancing encryption technologies \cite{[285],[286]} have emerged as an innovative solution, effectively addressing many challenges in data sharing. This technology enables specific computation on encrypted data. Users, by decrypting the data with their designated keys, can access solely the output of the specific function, thereby ensuring that other irrelevant information remains inaccessible \cite{[10]}.\par
Privacy-enhancing encryption technologies can address the data sharing dilemma from multiple dimensions: At the security level, Attribute-Based Encryption (ABE) \cite{[74],[104]} and Identity-Based Encryption (IBE) \cite{[11],[59]} enable authorization-based decryption according to user attributes or identities-ciphertext can only be decrypted when specific conditions are met, thereby accurately enhancing data sharing security. At the flexibility level, Searchable Encryption (SE) \cite{[18],[112],[58],[78]} supports the retrieval of shared data in cloud environments or distributed systems via keyword indexing, while Proxy Re-encryption (PRE) \cite{[7],[54],[99]} allows ciphertext re-encryption before transmission. This eliminates the cumbersome steps of downloading, decrypting, and retransmitting data in traditional sharing processes, facilitating cross-security-domain data flow. At the efficiency level, Broadcast Encryption (BE) \cite{[54],[28]} can achieve multi-user sharing through open channels and improve sharing efficiency, which is crucial for scenarios such as large-scale model training. It should be noted that improvements in security and functionality often come at the cost of partial performance, and balancing these three aspects remains a core scientific issue that urgently needs to be resolved. Current research primarily focuses on one or two key dimensions among them.\par
With the in-depth penetration of Artificial Intelligence (AI) technology, the risk of privacy leakage in intelligent scenarios is on the rise, and the tension between privacy protection and intelligent experience has become increasingly pronounced \cite{[309]}. In smart homes, the risk of exposure of sensitive information, such as sleep monitoring data and household images, is extremely high \cite{[305],[306]}. In the field of intelligent transportation, the leakage of users' travel trajectories may lead to security problems such as tracking \cite{[307],[308]}.
Notably, the personalized service capability of intelligent agents essentially hinges on user data collection and model training. For example, the collaboration of cross-room sensor data to achieve intelligent temperature control, and the in-depth mining of user travel data, form the basis for optimizing route planning. However, it is difficult to strike a balance between privacy and intelligent experience solely through data isolation or desensitization.
Therefore, the collaboration between privacy-protecting encryption algorithms and enhancement technologies (Federated Learning (FL), Secure Multi-Party Computation (MPC), and other relevant technologies are involved, for details, refer to Sec.\ref{2.2}) serves as the core solution to this contradiction. Firstly, encryption algorithms block privacy leakage channels at the source. Secondly, enhancement technologies support the joint modeling of multi-source data, thereby improving service accuracy while safeguarding privacy. Together, these two types of technologies can balance privacy protection and data value release, providing core technical support for the sustainable development of scenarios such as smart homes and intelligent transportation.\par
%Traditional encryption technologies can only ensure the security of data sharing. They struggle to meet the demands of today’s increasingly complex environment \cite{[286]}. The emergence of functional encryption \cite{[285],[286]} technology has provided effective solutions to many issues in data sharing \cite{[10]}. Functional encryption technology realizes specific calculation of data through encryption function. Users can decrypt encrypted data and obtain the output result of specific function according to the key they hold, but they cannot obtain other irrelevant information \cite{[286]}. 
%Jin et al. \cite{[246]} focused on blockchain-based secure and private data sharing. 

While prior surveys have extensively examined data sharing challenges in domains such as healthcare \cite{[251],[252],[253]}, agriculture \cite{[256]}, social networks \cite{[248]}, and e-governance \cite{[249],[250]}, they remain anchored in traditional security paradigms. The landscape shifted in 2020 when Keller introduced the seminal framework ``a versatile framework for MPC'' \cite{[303]}, heralding the large-scale deployment of privacy-preserving computation secure MPC. Subsequent research has grown exponentially, transforming the mantra of ``data available but invisible''  from aspiration to operational reality \cite{[304]}. Existing reviews, however, are constrained by their preoccupation with static data security, overlooking the paradigm shift catalyzed by emerging privacy-enhancing technologies \cite{[245],[256]}. Song et al. \cite{[247]} systematically survey 109 blockchain-based data sharing articles (2012-2021) and dissect their architectures, interoperability, and security mechanisms. Gupta et al. \cite{[243]} present a taxonomy of data-protection techniques spanning 49 articles (2010-2020), whereas Upadrista et al. \cite{[244]} synthesize 48 articles (2013-2022) on blockchain adoption in remote-health-monitoring scenarios, emphasizing data security and privacy. In contrast, this survey achieves threefold advances. First, it consolidates 116 core articles and incorporates an additional 48 top-tier conference and journal papers, published between 2023 and 2025, yielding the most comprehensive literature map to date. Second, it systematically catalogues emerging attacks across the entire data sharing life cycle. Third, it introduces a quantifiable security-performance-functionality evaluation matrix, enabling granular comparison of each proposed scheme. 
\par 

\begin{figure*}[!t]
    \centering
    \includegraphics[width=14cm]{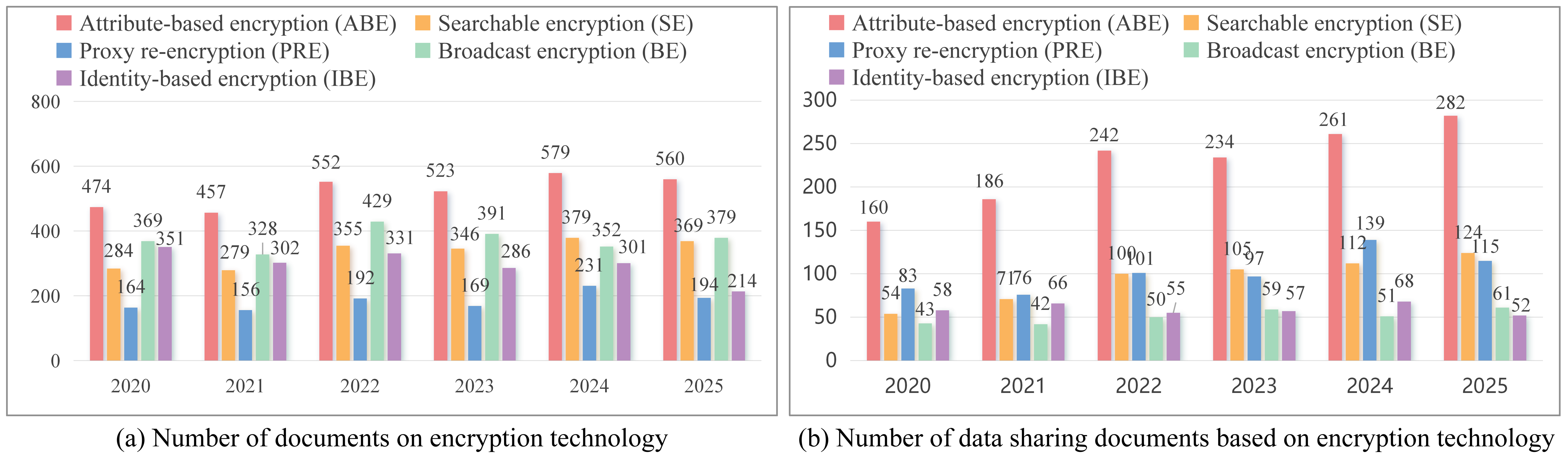}
    \caption{Research trends on privacy-enhancing encryption technologies and data sharing}
    \label{p1}
\end{figure*}

From 2020 to 2025, this work conducted a statistical analysis of literature related to privacy-enhancing encryption technologies and data sharing based on privacy-enhancing encryption. The analysis covered journal papers, conference papers, patents, and dissertations. The results are presented in Fig.\ref{p1}, respectively. Comprehensive analysis reveals that research based on ABE, SE and PRE exhibits a continuous growth trend. In contrast, research on BE and IBE fluctuates but maintains a relatively stable overall number. Given this, this work will center on applying ABE, SE, and PRE to data sharing. We'll systematically analyze 116 related papers and delve into their multi-dimensional impacts on security, performance, and functional adaptability. SE and IBE are incorporated into the enhancement technology. The contributions of this work are as follows:\par
%Through an in-depth analysis of related research, this paper attempts to address the following key questions:

\begin{itemize}
\item[1)] Three privacy-enhancing encryption technologies, ABE, SE and PRE, are systematically summarized and classified. 12 practical enhancement technologies in the data sharing scheme are summarized.\par
\item[2)]Through a meticulous review of the entire data sharing process, 20 potential attacks across different data sharing stages are deeply analyzed. This analysis offers precise targets for enhancing data sharing security.\par
%This paper pioneers a clear and universal data sharing process framework. It ensures the systematization and standardization of data sharing in all directions. By meticulously combing through the entire data sharing process, it deeply summarizes and analyzes 17 potential attack types across different data sharing stages. This provides precise targets for data sharing security protection.
\item[3)] The crucial roles of combining privacy-enhancing encryption with enhancement technologies in bolstering data sharing security, optimizing performance, and broadening functional applications are comprehensively compared and analyzed. Additionally, the paper explores how existing research balances security, functionality, and performance in data sharing.\par
%This paper investigates 11 enhancement technologies and the crucial roles of  ABE, SE, and PRE combined with enhancements. It compares how these technologies boost data sharing security, optimize performance, and expand functionality. Additionally, it offers a meticulous analysis of their unique advantages.
\item[4)] The paper systematically outlines optimal solutions for privacy-enhancing encryption based data sharing across various scenarios. It also sums up and examines key challenges in current research, proposing future research directions in this field based on this analysis.
%This paper systematically sorts out typical application scenarios of data sharing based on functional encryption algorithms across multiple fields. It summarizes and analyzes key challenges in existing research and prospectively proposes future research directions in this field.
\end{itemize}\par
The structure of this work is as follows: Sec.\ref{s2} presents an in-depth overview of the principles and classification of ABE, SE, and PRE, while providing a detailed introduction to the practical enhancement technologies in data sharing schemes. Sec.\ref{s3} proposes a general data sharing process and analyzes the security threats faced by each stage. Sec.\ref{s4}, \ref{s5}, \ref{s6} respectively summarize the research progress of ABE, SE, and PRE in improving data sharing security, performance, and functionality. Sec.\ref{s7} discusses application scenarios of data sharing based on privacy-enhancing encryption technologies. Sec.\ref{s8} delves into this research, summarizes challenges, and looks forward to future research directions. Sec.\ref{s9} concludes the paper.\par

\section{Background}\label{s2}
This section provides a systematic overview of the principles and research status of ABE, SE, and PRE. It also summarizes the enhancement technologies that may be relevant to data sharing schemes.
\subsection{Principles and Research Status of ABE, SE and PRE}
This section expounds the basic contents and classification of ABE, SE and PRE. ABE implements access authorization management based on attributes through fine-grained permission control strategy, while SE focuses on efficient keyword retrieval under the condition of steganography, taking into account privacy protection and data availability. PRE, on the other hand, takes security permission conversion as the core to realize the controllable migration of encrypted data between different cryptographic systems. Table  \ref{table1} summarizes the main characteristics of ABE, SE and PRE.

%表1
\begin{table*}[ht]
\centering
\footnotesize 
\setlength{\tabcolsep}{4pt} % 调整列间距以适应双栏格式
\caption{Main characteristics of ABE, SE and PRE} % 添加表格标题\textbf{}
\label{table1}
\begin{tabular}{m{1.5cm} m{3.8cm} m{2.8cm} m{4.5cm}} % 调整最后一列宽度以更好地占满双栏
\hline
\textbf{Technology} & \textbf{Application Characteristics} & \textbf{Superiority} & \textbf{Limitations} \\
\hline
ABE & Frequent retrieval of encrypted data content is required. & Efficient query in encrypted environment. & Only keyword search is supported, not complex analysis. \\
\hline
SE & Need to entrust a third party to forward or share encrypted data. & Cross-domain data security flow. & Relying on trusted agents may introduce a single point of risk. \\
\hline
PRE & Need dynamic and fine-grained permission control. & Role/Attribute-Driven Access Control. & The calculation overhead is high, and policy management is complicated. \\
\hline
\end{tabular}
\end{table*}

\subsubsection{Attribute-Based Encryption}
\begin{figure*}[!t]
    \centering
    \includegraphics[width=14cm]{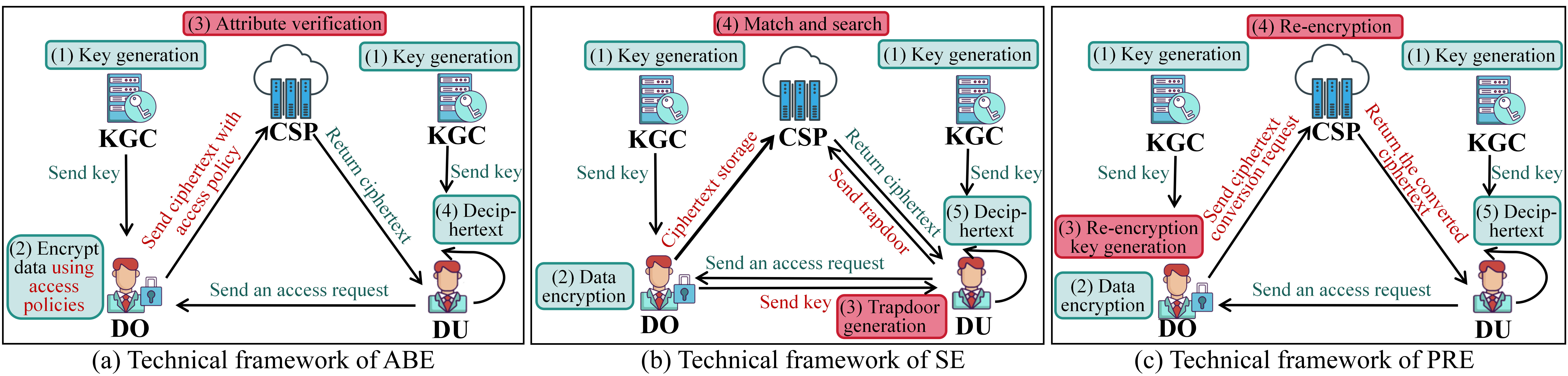}
    \caption{Technical framework of ABE, SE and PRE}
    \label{p3}
\end{figure*}

ABE is a fine-grained data access control encryption technology that links user attributes to access policies. Fig. \ref{p3} (a) illustrates the ABE technical framework. The detailed process is as follows: 1) Key Generation: Key Generation Center (KGC) generates keys for encryption and decryption, and sends them to Data Owner (DO) and Data User (DU). 2) Encrypt Data Using Access Policies: The DO encrypts the data with the key and access policy received from KGC to ensure that only users who meet the access policy can decrypt the data. 3) Attribute Verification: Cloud Server Provider (CSP) receives the encrypted data and access policy from DO, and verifies whether the attributes of DU conform to the access policy. 4) Decipher Text: DU decrypts the encrypted data received from CSP with the key received from KGC to obtain the original unencrypted data. \par
Based on where the policy is bound, it can be categorized into Ciphertext-Policy Attribute-Based Encryption (CP-ABE) \cite{[37],[50],[69],[93]} and Key-Policy Attribute-Based Encryption (KP-ABE) \cite{[27]}. CP-ABE embeds the access policy in the ciphertext and associates the user key with the attribute set. Data owners can define dynamic access rules, and only users meeting these policies can decrypt the data. This approach supports one-to-many sharing and suits scenarios where DO actively control access rights. KP-ABE, by contrast, embeds the access policy in the user key and links the ciphertext to the attribute set, making it fit for situations where the receiver passively matches the policy, such as in paid content distribution. \par
In terms of dynamic support, ABE can be divided into static attribute encryption \cite{[82]} and Dynamic Attribute-Based Encryption (DABE) \cite{[50],[74]}. In static attribute encryption, user attributes can't be altered after key generation, so it's suitable for scenarios with stable attributes. DABE, however, allows real-time attribute and policy updates. But it faces the key revocation problem.\par
Regarding privacy protection, ABE can be classified into hidden-policy attribute encryption \cite{[45],[50]} and anonymous attribute encryption \cite{[23]}. Hidden-policy attribute encryption conceals the access policy during encryption, such as anonymous access trees, to prevent servers or third parties from getting sensitive rule information. This makes it ideal for commercial bidding or sensitive policy protection. In anonymous attribute encryption, the user's key and attributes are anonymized to prevent user identity inference from the key. \par

\par
\subsubsection{Searchable Encryption}
SE is a cryptographic technology that facilitates keyword searches on encrypted data. Fig. \ref{p3} (b) illustrates the SE technical framework. The detailed process is as follows: 1) Key Generation: KGC generates keys for encryption and decryption, and sends them to DO and DU. 2) Data Encryption: DO encrypts the data with the key received from KGC, and sends the data ciphertext to CSP for storage. 3) Trapdoor Generation: DU generates a trap door, which is a key for searching encrypted data. DU sends this trap to CSP. 4) Match and Search: CSP receives the trap door from DU and uses this trap door to search in the encrypted data to find the matched encrypted data item. CSP sends the search results back to DU. 5) Decipher Text: DU decrypts the search result received from CSP using the key received from KGC to obtain the original data. \par 
Based on the cryptosystem classification, it can be categorized into Symmetric Searchable Encryption (SSE) \cite{[193]} and Asymmetric Searchable Encryption (ASE) \cite{[233]}. In SSE, encryption, decryption, and trapdoor generation all utilize the same symmetric key. It relies on technologies such as pseudo-random functions and hash algorithms to construct indexes and search protocols. SSE boasts high computational efficiency and a simple algorithm structure, making it suitable for single-user models. However, its flexibility is somewhat limited, and it typically only supports single-keyword or simple logical searches. ASE, on the other hand, is based on public-key cryptography, such as bilinear pairings and elliptic curves. DO encrypts the data using the public key, while the user generates the search trapdoor with the private key. This makes ASE suitable for multi-user sharing scenarios like email systems, as it eliminates the need for key negotiation. Nevertheless, it comes with a higher computational cost and is better suited for complex search logics, such as multi-keyword combinations.\par
According to keyword processing capabilities, SE can be divided into single-keyword search \cite{[232]} and multi-keyword search \cite{[77],[153]}. Single-keyword search only supports one keyword at a time. It is relatively simple to implement but has lower accuracy. Multi-keyword search, however, supports queries with multiple keyword combinations, such as ``AND''/``OR'' logic. This requires addressing challenges related to computational complexity and privacy leakage.\par
From a dynamic perspective, SE can be classified as Static Searchable Encryption \cite{[1],[102]} and Dynamic Searchable Encryption (DSE) \cite{[115],[159]}. Static Searchable Encryption cannot be modified once the data is uploaded, making it suitable for archiving scenarios. DSE, in contrast, supports data addition, deletion, and modification. It needs to address issues like forward privacy and backward privacy.\par
\subsubsection{Proxy Re-encryption}
PRE is an encryption technology that allows a third-party Proxy to transform encrypted data from one user's key system to another without decrypting the ciphertext. This technology elegantly solves the key management problem in encrypted data sharing Fig. \ref{p3} (c) illustrates the PRE technical framework. The detailed process is as follows: 1) Key Generation: KGC is responsible for generating initial encryption keys and sending them DO and DU. 2) Data Encryption: DO encrypts the data with the key received from KGC to ensure the security of the data during storage or transmission. 3) Re-encryption Key Generation: DO generates a re-encryption key, which is used to convert the data from DO's encryption domain to DU's encryption domain. DO sends this re-encryption key to CSP. 4) Re-encryption: CSP uses the re-encryption key received from DO to re-encrypt the encrypted data stored in the cloud, so that it can be converted to the encryption domain of DU without decryption. 5) Decrypt Ciphertext: DU decrypts the re-encrypted data received from CSP using the key received from KGC, thus obtaining the original data. \par 
Based on the number of conversion times, PRE can be categorized into single-hop PRE \cite{[6],[55],[73]} and multi-hop PRE \cite{[63],[57],[73],[87]}. Single-hop PRE permits the proxy to convert the ciphertext only once, and the converted ciphertext cannot be further processed by other proxies. This type of PRE is well-suited for single authorization scenarios. It offers high security but has limited flexibility. Multi-hop PRE, on the other hand, supports multiple conversions of the ciphertext by the proxy. After each transformation, the ciphertext can still be further processed by other proxies. This type of PRE also has high security and is suitable for single authorization scenarios.\par
According to the direction of authorization, PRE can be divided into unidirectional PRE \cite{[6],[63]} and bidirectional PRE \cite{[273]}. Unidirectional PRE only allows one-way conversion from the licensor to the licensee. The reverse operation is not feasible. This type of scheme is common in cloud data sharing. Bidirectional PRE supports two-way authority conversion. In other words, user A and user B can authorize decryption authority to each other. It is suitable for peer-to-peer cooperation scenarios, but requires a higher-level security verification mechanism.\par
Based on the fundamental classification of cryptography, PRE can be divided into bilinear pair-based PRE \cite{[54]} and lattice-based PRE \cite{[16]}. Bilinear pair-based PRE leverages bilinear pairings on elliptic curves to achieve ciphertext conversion. This type of PRE has the advantage of high computational efficiency, but it depends on specific mathematical assumptions. Lattice-based PRE is based on quantum-resistant lattice cryptography. It is suitable for post-quantum security scenarios.

\subsection{Principle and Characteristics of Enhancement Technology}

\label{2.2}
%In the realm of data sharing, encryption technology undoubtedly offers a fundamental assurance of data confidentiality. However, practical applications are still beset by a array of challenges. For instance, the risk of privacy disclosure can't be entirely precluded. Besides, there's a lack of flexibility in data sharing, and the computational overhead associated with encryption algorithms can substantially impinge on sharing efficiency. To address these extant issues, this paper classifies the technical approaches in data sharing schemes into three major categories: security improvement, performance improvement, and function improvement. Security improvement centers on counteracting various types of attacks and bolstering privacy protection. Performance improvement is committed to optimizing the computational complexity and resource consumption of the scheme. Function improvement is dedicated to broadening the scope of shared-scenario functions. This includes supporting joint computing and enabling fine grained permission control. The following will explain the improvement path of some enhancement technologies to data sharing from the technical dimension.
In data sharing scenarios, while encryption technology ensures basic data confidentiality, practical applications still encounter multi-dimensional challenges. This section introduces 12 practical enhancement technologies in data sharing schemes.\par 
%\Lv{Shorten the introduction of each technology.}
\begin{itemize}
    \item[1)] Cloud Computing (CC): CC \cite{[105],[27],[19]} provides computing resources distributed on demand through the network. Users can flexibly acquire and release resources and pay according to usage without managing the underlying hardware. 
%It includes the following key technologies: 1) Virtualization Technology: Through virtualization technology, physical servers, storage devices, and network resources are abstracted into virtual resource pools, which improves resource utilization and flexibility. 2) Storage technology: Including distributed storage and cloud storage to ensure high availability, reliability, and security of data. 3) Automated management: Through automated tools and platforms, resources can be deployed, monitored, and managed quickly, and manual intervention can be reduced.

    \item[2)] Blockchain (BC): BC \cite {[1],[14],[301]} is a multi-party maintained distributed ledger technology that leverages cryptography for data transmission and access security, enabling tamper-resistant, non-repudiable, and consistent data storage.
    %It records transactions through sequentially connected data blocks to ensure data integrity and non-tampering.
%and it includes the following key technologies: 1) Distributed Ledger Technology (DLT): The core of blockchain is distributed account book. All participants jointly maintain an account book, and each node keeps a complete copy of the account book to ensure the transparency and consistency of data. 2) Cryptography technology: Including hash algorithm and asymmetric encryption to ensure the integrity of data and the security of transactions. 3) Consensus Mechanism: Used to reach consensus in distributed networks. Common consensus algorithms include Proof of Work (PoW), Proof of Stake (PoS), Practical Byzantine Fault Tolerance (PBFT), and so on. 4) Smart Contract: An automatically executed contract clause, which is deployed in the blockchain in the form of code and automatically executed when certain conditions are met, without the intervention of a third party. 5) Peer-to-peer Networking (P2P): The blockchain runs on a decentralized P2P network, and each node is both a data storage and a data verifier.

    \item[3)] Interplanetary File System (IPFS): IPFS \cite{[24],[35]} is a P2P distributed file storage and sharing protocol. Through content addressing and distributed storage, it aims to create a more efficient, secure, and lasting Internet file storage and transmission system, with the goal of supplementing or even replacing HTTP and improving data availability and security. 

    \item[4)] Zero-Knowledge Proof (ZKP): ZKP \cite{[237]} is a cryptographic protocol. The prover can prove the truth of a statement to the verifier without revealing any additional information. The verifier can only confirm whether the statement is true or not. 
%It has the following three main features: 1) Completeness: If the statement is true and the prover follows the protocol, the verifier will be persuaded with high probability. 2) Soundness: If the statement is false, no deceptive prover can convince the verifier (unless the probability is extremely small). 3) Zero-Knowledge: The verifier cannot obtain any information beyond the validity of the statement from the proof process.

    \item[5)] Homomorphic Encryption (HE): HE \cite{[6]} is an encryption technology, which allows direct calculation on encrypted data without decryption first. The decrypted calculation results are the same as those directly calculated in plain text, ensuring data privacy. 
%HE has the following classification: 1) Additive Homomorphic Encryption (AHE): Allows the addition of encrypted data. For example, the Paillier encryption algorithm supports addition and multiplication of ciphertext. 2) Multiplicative Homomorphic Encryption (MHE): Allows multiplication of encrypted data. For example, the Rivest-Shamir-Adleman (RSA) encryption algorithm supports multiplicative homomorphism under some conditions. 3) Fully Homomorphic Encryption (FHE): Supports any complicated calculation of encrypted data, including the combination of addition and multiplication.

    \item[6)] Secure Multi-party Computing: MPC \cite{[162]} is a cryptographic protocol, which allows multiple parties to jointly calculate function results without a trusted third party, ensuring the privacy of input data, only outputting the results and protecting the privacy of all parties. 
%It includes the following key technologies: 1) Secret Sharing: A secret is divided into multiple parts and distributed to different participants. Only when enough participants cooperate can the original secret be recovered. 2) Garbled Circuit: The calculation process is compiled into a Boolean circuit, and the truth table of the circuit is encrypted and scrambled, so that the participants can complete the calculation without leaking the input. 3) Oblivious Transfer (OT): A protocol that allows one party to obtain information from another party, but the sender does not know what information the receiver has obtained. For example, Alice has two secrets, and Bob can choose one of them, but Alice doesn't know which one Bob chose. 
%4) HE: Allows direct calculation on encrypted data, and the calculated results are the same as those calculated in plain text after decryption. 5) ZKP: Used to verify the correctness of calculation results without revealing any additional information.

    \item[7)] Differential Privacy (DP): DP \cite{[123]} is a privacy protection technology, which ensures the privacy of data when it is released by adding noise to data analysis. Its core is that the output distribution of adjacent data sets is almost the same, even if the attacker has a lot of background knowledge, it is impossible to infer the privacy information of a single user. 
%It includes the following key technologies: 1) Randomized Response: A local differential privacy technology, in which users randomly disturb the data locally before uploading. 2) Privacy Budget Management: Balance privacy protection and data availability in different queries by reasonably allocating privacy budget.

    \item[8)] Federated Learning: FL \cite {[162],[319]} is a distributed machine learning technology that supports multi-party joint model training without sharing original data. Participants process data and update parameters locally, then upload parameters to a central server for aggregation, thus protecting data privacy and achieving collaborative modeling.
%It includes the following key technologies: 1) Distributed Training: Federated learning adopts a distributed training framework, and multiple clients (such as mobile devices, Internet of Things devices, etc.) process data and update models locally. 2) Model aggregation algorithm: The updated model parameters of each client are fused by an aggregation algorithm to generate a global model. Common algorithms include Federated Averaging (FedAvg), which aggregates the model parameters of clients through weighted averaging. 3) Privacy protection mechanism: Federated learning avoids the sharing of original data by processing data locally, thus protecting users' privacy. Combined with DP technology, privacy protection is further enhanced, such as adding noise during model updating.

    \item[9)] Signcryption: Signcryption \cite{[137]} combines digital signature and encryption functions to realize data confidentiality, integrity, and identity authentication. It encrypts data by unified operation and signature verification, which is more efficient and safer than using signature and encryption alone, and is suitable for various security requirements scenarios. 
%It includes the following key technologies: 1) Combination of encryption and signature: Signcryption integrates encryption and signature operations through a specific algorithm, avoiding the complexity and efficiency problems of separate encryption and signature. 2) Design of security protocol: The signcryption needs to design a secure protocol to ensure that the confidentiality, integrity, and identity authentication of data will not be affected under various network environments and attack scenarios.

%It includes the following key technologies: 1) Content Addressing: IPFS uses the hash value of the content (such as SHA-256) as the unique identifier of the file, instead of the URL based on the storage location of the file. This ensures that the contents of the file are unique and cannot be tampered with. 2) Distributed Hash Table (DHT): IPFS uses DHT to store and retrieve the mapping relationship between file hash values and node addresses, and quickly locate the nodes where files are stored. 3) P2P Network and NAT Traversal: IPFS is based on a P2P network architecture, and node communication in different network environments is realized through the Interactive Connectivity Establishment Network Address traversal framework.
%(integrating Session Traversal Utilities for NAT (STUN), Traversal Using Relays around NAT (TURN), and other protocols)
%\cite{[173],[183],[148],[140],[195]}
    \item[10)] Identity-Based Encryption (IBE): IBE \cite{[73],[86],[11]} is a public key encryption technology. The public key of users is directly related to their identity information, and the private key is generated and distributed by a trusted KGC. This mechanism simplifies public key management and makes encryption and decryption efficient and convenient. 
%It includes the following key technologies: 1) Pairing technology: Pairing technology is an important mathematical tool to realize IBE, which allows special operations between points on elliptic curves. For example, the Boneh-Franklin scheme uses elliptic curve pairing to realize IBE. 2) Binding of identity and public key: The user's public key is generated directly from his identity information (such as e-mail address) and the system's master public key, without complicated public key distribution like traditional public key encryption. 3) Security mechanism: In order to prevent PKG from abusing its power, IBE systems usually introduce additional security mechanisms, such as key splitting and authentication.
%\cite{[152],[224],[186],[227]}
    \item[11)] Broadcast Encryption (BE): BE \cite{[95],[201]} allows broadcast centers to efficiently transmit encrypted information to multiple users, and only authorized users can decrypt it. It flexibly selects authorized users to prevent unauthorized access, and is widely used in digital rights management, satellite communication, and multimedia broadcasting. 
    \item[12)] Post-Quantum Cryptography (PQC): PQC \cite{[320],[322],[323]} is a set of cryptographic algorithms and protocols specially designed to resist the attacks of quantum computers. Its core goal is to ensure the confidentiality, integrity and availability of information transmission and storage in the era of quantum computing.
%It includes the following key technologies: 1) Key management: Broadcast encryption requires effective key management to ensure that authorized users can obtain decryption keys, while unauthorized users cannot. Key management usually involves a Key Distribution Center (KDC) or a similar mechanism. 2) Dynamic user management: Broadcast encryption supports adding and deleting users dynamically, and ensures the flexibility and security of the system by updating the key or adjusting the subset coverage strategy.
\end{itemize}

\section{Data Sharing Process and Its Security Threats}\label{s3}
This section begins by elaborating on the universal data sharing process framework proposed in this work. It then thoroughly summarizes and analyzes 20 potential attack types that can occur across different stages of data sharing. This analysis offers security protection targets for data sharing processes.

\subsection{Data Sharing Process Framework}
\begin{figure*}[!t]
    \centering
    \includegraphics[width=13.8cm]{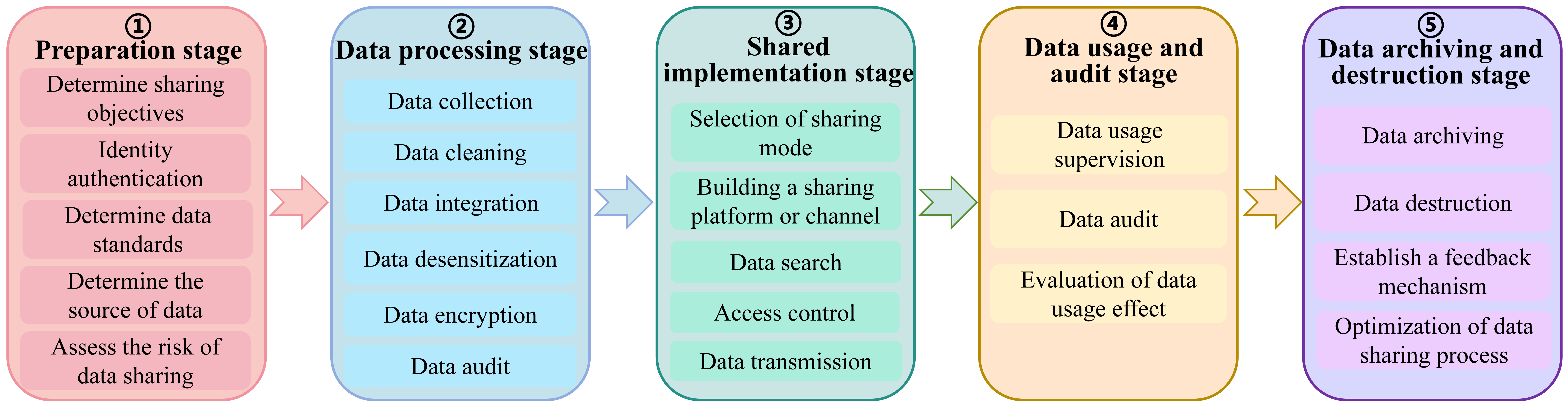}
    \caption{Data sharing process framework}
    \label{p2}
\end{figure*}

this work constructs a data sharing process framework that fully considers security, efficiency, and sustainable development. When devising this process, we first thoroughly study by multinational data security regulations \cite{[277],[284]}, industry regulatory policies and government documents \cite{[280]} to guarantee the legality and compliance of the data sharing process. Meanwhile, we actively draw on industry standards \cite{[281],[282]}, international norms \cite{[278]}, and best practices \cite{[283]} to ensure the process is scientific and advanced. Notably, specific industries can further refine the data sharing operation mode according to their business needs, organizational structure, and technical capabilities, all within this process framework. For details of the data sharing process framework constructed in this work, refer to Fig.\ref{p2}. The following provides a detailed description of each stage of the process.
%The data sharing process constructed in this paper is formed on the basis of all-round consideration of data sharing security, efficiency and sustainable development. In the process of formulation, firstly, the multinational data security regulations, industry regulatory policies and government documents are thoroughly studied and strictly followed to ensure the legal compliance of data sharing process. At the same time, actively learn from industry standards, international norms and best practice experience to ensure the scientific and advanced process. In particular, specific industries can further refine the specific operation mode of data sharing according to their own business needs, organizational structure and technical capabilities under the framework of this process. See Fig.\ref{p2} for details of the data sharing process framework constructed in this paper. Each stage of the process will be described in detail below.

\subsubsection{The Preparation Stage}
The initial stage of data sharing is the preparation stage of sharing. At this stage, it is necessary to clarify the data standards and access rights, that is, to determine who can obtain what data. The specific contents include: defining the sharing goal and demand, authenticating the data demander, determining the data standard and source, and evaluating the risk of data sharing.

\begin{itemize}
    \item Determine sharing objectives and requirements: Define the objectives, uses and expected results of data sharing, and determine the types and scope of data to be shared.
    \item Identity authentication: Strictly authenticate and authorize the identity of the data demander, and only licensed users who pass the authentication can proceed to the next stage.
    \item Determine data standards: Specify data format, fields, value range, etc. to ensure data consistency and availability.
    \item Determine the source of data: The source of data determines the quality and reliability of data. Data can come from public databases, professional institutions, internal data warehouses of enterprises, etc., and it is necessary to ensure the legitimacy and reliability of data sources.
    \item Assess the risk of data sharing: Conduct a comprehensive risk assessment of the shared data, including data leakage risk, privacy invasion risk and data quality risk, and formulate corresponding risk response measures.
\end{itemize}

\subsubsection{Data Processing Stage}
The data processing stage is the key link before data sharing. In this stage, a series of operations such as data collection, cleaning, integration, desensitization, encryption and audit are carried out in turn to ensure the confidentiality, integrity and availability of data in all directions.

\begin{itemize}
    \item Data collection: Collect relevant data according to the determined data source and collection method. In the process of collection, we should pay attention to protecting the privacy and security of data and abide by relevant laws and regulations.
    \item Data cleaning: Preprocess the collected data, remove redundant data, correct errors, fill in missing values, etc., so as to improve the quality and usability of data.
    \item Data integration: Data from various sources are integrated according to the determined data standards, so as to solve the problems of inconsistent data formats, fields and value ranges, and form a unified data set, which provides a basis for subsequent data analysis and sharing.
    \item Data desensitization: Desensitize sensitive data and remove the identification that can identify personal identity or sensitive information.
    \item Data encryption: Data is encrypted by encryption technology to prevent data leakage and unauthorized access, and to protect data privacy and security.
    \item Data audit: Before data sharing, conduct strict quality audit on data to ensure consistency, integrity and accuracy of data, ensure credibility and effectiveness of data, and reduce potential risks caused by quality problems.
\end{itemize}

\subsubsection{Shared Implementation Stage}
In the implementation stage of sharing, firstly, we need to select the sharing mode and build a data sharing platform or channel. In the process of sharing, access control is implemented, and fine-grained authorization is given to the data demander, and finally the data is delivered to the demander safely.

\begin{itemize}
    \item Selection of sharing mode: According to the nature and scale of data and the needs of participants, choose the appropriate data sharing mode, such as data interface sharing and data platform sharing.
    \item Building a sharing platform or channel: If it is a platform-based sharing mode, it is necessary to build corresponding data sharing platforms, such as data warehouses and data lakes, and configure related functions such as data storage, management and access rights. For data interface sharing, it is necessary to develop and deploy standardized interfaces.
    \item Data search: Many existing data are encrypted and stored in third-party platforms such as IPFS or cloud server. It is necessary to ensure that only authorized personnel can search relevant data, and at the same time ensure the security and privacy of the search process.
    \item Access control: Implement fine-grained access control measures, limit the scope, frequency and mode of data use, monitor and manage the behavior of visitors, and ensure the security and compliance of data sharing.
    \item Data transmission: Transmit prepared data to data demanders through the selected sharing mode, with guaranteed data security and integrity in transmission and storage.
\end{itemize}

\subsubsection{Data Usage and Audit Stage}
After the data demander obtains the data, it must strictly supervise and audit the data use process and evaluate its use effect.

\begin{itemize}
    \item Data usage supervision: The data usage process can be monitored and recorded in real-time, covering operations such as data access, downloading and modification, as well as information such as data usage time, place and personnel, so as to find and stop violations in time.
    \item Data audit: Audit the data usage regularly, check the compliance and safety of data usage, and make timely rectification when problems are found.
    \item Evaluation of data usage effect: Continuously monitor data usage and evaluate the effect of data sharing, including data usage frequency, usage scenarios, contribution to business, etc., so as to find problems in time and make optimization and adjustment.
\end{itemize}

\subsubsection{Data Archiving and Destruction Stage}
After the data is used, it is necessary to archive or destroy the data, and establish an effective feedback mechanism to optimize the data sharing process according to the feedback information.

\begin{itemize}
    \item Data archiving: Archive some data that are valuable for reference or need to be preserved for a long time in a specific storage medium and manage it properly so that it can be retrieved and restored when needed in the future.
    \item Data destruction: When the data is no longer needed, the data shall be destroyed according to the data destruction strategy to ensure that the data cannot be recovered or accessed, so as to prevent data leakage and abuse.
    \item Establish a feedback mechanism: Establish an effective feedback mechanism, collect data and put forward opinions and suggestions on the process of data sharing.
    \item Optimization of data sharing process: According to the feedback information, adjust and improve the strategy, process and technical means of data sharing to improve the efficiency and effect of data sharing.
\end{itemize}

\subsection{Security Threats to The Data Sharing Process}
There are many possible attacks in the process of data sharing. Table \ref{table5} shows the possible attacks in each stage of each data sharing process. The following are 20 common attacks summarized: 
%\Lv{Several new attacks have been added, and references have been added to each attack.}

\begin{table}[htbp]
\centering
\caption{Possible Security Attacks on Data Sharing Process}
%\feng{any citations?}
\label{table5}
\footnotesize
\begin{tabular}{p{0.25\linewidth} p{0.68\linewidth}}
\toprule
\textbf{Stage} & \textbf{Possible Security Attacks} \\
\midrule
Preparatory stage & Malware attack, Phishing attack, Spoofing attack, Insider attack, SCA, KGA, Quantum attack. \\
Data processing stage & Malware attack, DoS attack, HX-DoS attack, SQL injection attack, XSS attack, Zero-day attack, Eavesdropping attacks, Spoofing attack, Insider attack, Tampering attack, Forgery attack, ASA, Quantum attack. \\
Shared implementation phase & DoS attack, HX-DoS attack, MITM attack, SQL injection attack, XSS attack, Zero-day attack, Eavesdropping attacks, Spoofing attack, Replay attack, Hijacking attack, Tampering attack, Leak abuse attack, ASA, KGA, Quantum attack. \\
Data usage and audit & Malware attack, MITM attack, SQL injection attack, XSS attack, Zero-day attack, Spoofing attack, Insider attack, SCA, Replay attack, Hijacking attack, Tampering attack, Leak abuse attack, Forgery attack, Quantum attack. \\
Data archiving and destruction & Malware attack, DoS attack, HX-DoS attack, Spoofing attack, Insider attack, Tampering attack, Leak abuse attack. \\
\bottomrule
\end{tabular}
\end{table}

%\begin{enumerate}
\begin{itemize}
    \item[1)] Malware attack \cite{[292]}: The attacker uses malware to infect the target system, steal data, destroy the system or control equipment.
    \item[2)] Phishing attack \cite{[293]}: Attackers trick users into entering personal information by forging legitimate websites or sending false emails, thus stealing users' privacy.
    \item[3)] Denial of Service (DoS) attack \cite{[244]}: The attacker paralyzes the service of the data sharing platform or system through a large number of meaningless requests or data traffic, resulting in normal users unable to access data or use services.
    \item[4)] Hybrid Denial of Service (HX-DoS) attack \cite{[292]}: Attackers make the target system unable to respond to legitimate users' requests by various means, which eventually leads to the unavailability of services.
   % \item Man-in-the-middle attack: The attacker eavesdrops and tampers between the two communication parties to obtain communication content or tamper with communication data.
    \item[5)] Structured Query Language (SQL) injection attack \cite{[299]}: Tampering with or stealing database information by injecting malicious SQL statements into input fields.
    \item[6)] Cross-site scripting (XSS) attack \cite{[294]}: An attacker injects malicious scripts into a webpage, and when other users visit the webpage, the scripts will be executed in the user's browser, thereby stealing the user's session information or performing other malicious operations.
    \item[7)] Zero-day attack \cite{[295]}: Attacking by using software vulnerabilities that have not been discovered or repaired. Without prior preventive measures, such attacks can often cause serious damage.
    \item[8)] Eavesdropping attacks \cite{[292]}: Attackers illegally obtain sensitive information in transmission by listening to network communication.
    %Voucher stealing attack: The attacker obtains the user's login credentials through various means, thus impersonating a legitimate user to access the system and data.
    \item[9)] Spoofing Attack \cite{[300]}: Attackers cheat the system or users by pretending to be legitimate users or devices and sending false information or data, thus gaining unauthorized access or sensitive information.
    \item[10)] Insider attack \cite{[292]}: Internal personnel may leak or tamper with data due to negligence or malicious behavior.
    \item[11)] Side channel attack (SCA) \cite{[292]}: Sensitive information is obtained by analyzing indirect information generated by the system during operation, such as power consumption and electromagnetic radiation.
    %\item Near-source attack: The attacker directly contacts the target system or network through physical contact or infiltration in the local area network to attack.
    %Springboard attack: The attacker first invades a low-security account or resource in the cloud service, and then uses it as a springboard to further attack other high-value targets.
    \item[12)] Man-in-the-middle (MITM) attack \cite{[244]}: An attacker may intercept data, steal sensitive information or tamper with data content during data transmission.
    \item[13)] Replay attack \cite{[292]}: The attacker intercepts and retransmits the previously legal data transmission to cheat the system or gain improper access rights, such as replaying the authentication request to gain continuous access.
    \item[14)] Hijacking attack \cite{[292]}: The attacker hijacks resources in the process of data sharing, such as network bandwidth, computing resources, etc., through malicious software or network attacks, which affects the normal data sharing.
    \item[15)] Tampering attack \cite{[292]}: The attacker falsifies the usage of data after sharing, such as modifying data usage records and audit logs to cover up illegal data usage.
    %\item Data abuse attack: The data demander may violate the data sharing agreement and use the shared data for unauthorized purposes, such as selling the data to a third party and using it for commercial competition.
    \item[16)] Forgery attack \cite{[300]}: Malicious users may generate false data, which will affect the quality and authenticity of data and lead to subsequent analysis and decision-making errors.
   \item[17)] Leak abuse attack \cite{[290]}: The attacker uses the specific information leaked by the encryption system and his prior knowledge to infer sensitive information, such as query content or data plaintext.
   \item[18)] Algorithm substitution attack (ASA) \cite{[302]}: Attackers use malicious algorithms with back doors to replace normal encryption, signature and other cryptographic algorithms, so that users can reveal secret information unconsciously.
   \item[19)] Keyword guessing attack (KGA) \cite{[77]}: Attackers gain unauthorized access to systems or data by guessing or using automated tools to try common keywords (such as passwords, keys, sensitive data, etc.).
   \item[20)] Quantum attack \cite{[59]}: Attackers use the powerful computing power of quantum computing to crack traditional encryption algorithms or implement other network security threats.
    %\item Data leakage attack: The attacker obtains sensitive data of enterprises or individuals through illegal means, causing serious economic losses and reputation damage.
\end{itemize}

\section{Security Improvement of Data Sharing}\label{s4}
As digital transformation speeds up, numerous security challenges emerge during data sharing. This section explores how ABE, SE, and PRE bolster data sharing security. It systematically sums up and analyzes relevant studies, as presented in Table \ref{table2}. 
%\Lv{Reorder according to the order of classification and appearance}

% 定义空心圆
\newcommand{\0}{%
    \begin{tikzpicture}[scale=0.1]
        \draw (0,0) circle (1cm);
    \end{tikzpicture}%
}

% 定义填充三分之一的圆
\newcommand{\1}{%
    \begin{tikzpicture}[scale=0.1]
        \draw (0,0) circle (1cm);
        \fill[black] (0,0) -- (0:1cm) arc (0:120:1cm) -- cycle;
    \end{tikzpicture}%
}

% 定义填充三分之二的圆
\newcommand{\2}{%
    \begin{tikzpicture}[scale=0.1]
        \draw (0,0) circle (1cm);
        \fill[black] (0,0) -- (0:1cm) arc (0:240:1cm) -- cycle;
    \end{tikzpicture}%
}

% 定义填满的圆
\newcommand{\3}{%
    \begin{tikzpicture}[scale=0.1]
        \fill[black] (0,0) circle (1cm);
    \end{tikzpicture}%
}

\renewcommand{\arraystretch}{0.1}
\begin{longtable}{
>{\raggedright\arraybackslash}p{0.6cm} 
>{\raggedright\arraybackslash}p{5.1cm} 
>{\raggedright\arraybackslash}p{1.6cm} 
>{\raggedright\arraybackslash}p{2cm}
>{\centering\arraybackslash}p{0.7cm} 
>{\centering\arraybackslash}p{0.9cm}
>{\centering\arraybackslash}p{0.7cm}}
\caption{Summary of Research on Improving the Security of Data Sharing by ABE, SE and PRE} 
%\feng{why not categorize them by ABE/SE/PRE and put lines between each category to match with the following 3 sections?}
\label{table2}  \\
\hline
\footnotesize \textbf{Paper} & 
\footnotesize \textbf{Research Contents} & 
\footnotesize \textbf{Encryption Technology} & 
\footnotesize \textbf{Enhancement Technology}& 
\footnotesize \textbf{Secu\-rity} & 
\footnotesize \textbf{Perfor\-mance} & 
\footnotesize \textbf{Functi\-onality}  \\ %

\hline
\endhead 
\addlinespace[2pt] %
\hline
\endfoot 
\hline
\addlinespace[2pt] %
%\multicolumn{7}{l}{\footnotesize Note: The intensity from strong to weak is expressed as: \3 > \2 > \1.} 
\multicolumn{7}{l}{\footnotesize Note: The comparison of security strength,  performance and functionality is expressed as follows: \3 > \2 > \1. } 
%, indicating that variable \3 has the strongest influence on the dependent variable

\endlastfoot  
\addlinespace[2pt] %
\footnotesize\cite{[9]}&\footnotesize Distributed or multi-attribute authority&\footnotesize ABE, SE&\footnotesize BC&\2&\1&\2\\
\footnotesize\cite{[42]}&\footnotesize Distributed or multi-attribute authority&\footnotesize ABE, PRE&\footnotesize BC, IPFS&\2&\1&\2\\
\footnotesize\cite{[72]}&\footnotesize Distributed or multi-attribute authority&\footnotesize ABE&\footnotesize BC, ZKP, CC&\3&\2&\2\\
\footnotesize\cite{[80]}&\footnotesize Distributed or multi-attribute authority&\footnotesize ABE&\footnotesize ZKP, BC&\3&\--&\2\\
\footnotesize\cite{[93]}&\footnotesize Distributed or multi-attribute authority&\footnotesize ABE&\footnotesize BC&\3&\1&\1\\
\footnotesize \cite{[4]}&\footnotesize Distributed or multi-attribute authority&\footnotesize ABE&\footnotesize BC, CC&\3&\2&\1\\
\footnotesize\cite{[7]}&\footnotesize Distributed or multi-attribute authority&\footnotesize ABE,  PRE&\footnotesize BC, CC&\3&\1&\--\\
\footnotesize\cite{[33]}&\footnotesize Distributed or multi-attribute authority&\footnotesize ABE, PRE&\footnotesize BC, CC&\3&\1&\1\\
\footnotesize \cite{[82]} & \footnotesize Hiding strategy & \footnotesize ABE & \footnotesize BC & \2 & \2 & \2 \\
\footnotesize\cite{[14]}&\footnotesize Hiding strategy&\footnotesize ABE&\footnotesize BC&\3&\2&\1\\
\footnotesize\cite{[45]} & \footnotesize Hiding strategy & \footnotesize ABE & \footnotesize BC, CC & \3 & \-- & \2 \\
\footnotesize\cite{[50]}&\footnotesize Hiding strategy&\footnotesize ABE&\footnotesize BC, CC&\2&\2&\--\\
\footnotesize\cite{[49]}&\footnotesize Hiding strategy&\footnotesize ABE, SE&\footnotesize BC&\3&\1&\2\\
\footnotesize\cite{[323]}&\footnotesize Hiding strategy&\footnotesize ABE&\footnotesize PQC&\3&\2&\1\\
\footnotesize\cite{[23]} & \footnotesize Anonymous ABE & \footnotesize ABE & \footnotesize BC & \2 & \1 & \2 \\
\footnotesize\cite{[44]}&\footnotesize Anonymous ABE&\footnotesize ABE, SE&\footnotesize CC, BC, IPFS&\3&\2&\1\\
\footnotesize\cite{[100]} & \footnotesize Anonymous ABE & \footnotesize ABE & \footnotesize BC & \2 & \1 & \2 \\
%\footnotesize\cite{[110]} & \footnotesize Anonymous ABE & \footnotesize ABE & \footnotesize BC, CC & \2 & \1 & \2 \\
\footnotesize\cite{[102]}&\footnotesize Combined blockchain&\footnotesize SE&\footnotesize BC&\3&\2&\1\\
\footnotesize\cite{[18]}&\footnotesize Combined blockchain&\footnotesize SE&\footnotesize BC, CC&\3&\1&\1\\
\footnotesize\cite{[70]} & \footnotesize Combined blockchain & \footnotesize SE, PRE & \footnotesize BC & \2 & \1 & \2 \\
\footnotesize\cite{[25]}&\footnotesize Combined blockchain&\footnotesize SE, PRE&\footnotesize BC, CC&\3&\1&\1\\
\footnotesize\cite{[162]}&\footnotesize Privacy computing technology&\footnotesize SE&\footnotesize MPC, FL&\3&\2&\1\\
\footnotesize\cite{[79]} & \footnotesize Privacy computing technology & \footnotesize SE & \footnotesize - & \2 & \2 & \2 \\
\footnotesize\cite{[123]}&\footnotesize Privacy computing technology&\footnotesize SE&\footnotesize FL, DP&\3&\1&\1\\
\footnotesize \cite{[15]} & \footnotesize Privacy computing technology & \footnotesize SE & \footnotesize BC, CC & \2 & \2 & \2 \\
%\footnotesize\cite{[144]}&\footnotesize Privacy computing technology&\footnotesize SE&\footnotesize ZKP, BC&\3&\1&\1\\
%\footnotesize\cite{[133]}&\footnotesize Privacy computing technology&\footnotesize SE&\footnotesize BC, FL&\3&\2&\1\\
\footnotesize \cite{[2]} &\footnotesize Combined with cryptography knowledge & \footnotesize SE & \footnotesize CC & \3 & \1 &\1 \\
\footnotesize\cite{[38]}&\footnotesize Combined with cryptography knowledge&\footnotesize SE, ABE&\footnotesize BC, CC&\3&\2&\1\\
\footnotesize\cite{[59]}&\footnotesize Combined with cryptography knowledge&\footnotesize SE, PRE&\footnotesize CC, IBE, PQC&\3&\1&\1\\
\footnotesize\cite{[117]}&\footnotesize Combined with cryptography knowledge&\footnotesize SE, ABE, PRE&\footnotesize BC, IPFS&\2&\1&\1\\
\footnotesize\cite{[106]}&\footnotesize Combined with cryptography knowledge&\footnotesize SE&\footnotesize CC, IBE&\3&\1&\1\\
\footnotesize\cite{[62]}&\footnotesize Blockchain and cryptography knowledge&\footnotesize PRE&\footnotesize BC&\2&\1&\2\\
\footnotesize\cite{[103]}&\footnotesize Combined blockchain&\footnotesize PRE&\footnotesize BC, ZKP&\3&\1&\1\\
\footnotesize\cite{[111]}&\footnotesize Combined blockchain and HE&\footnotesize PRE&\footnotesize CC, HE&\3&\1&\1\\
%\footnotesize\cite{[60]}&\footnotesize Combined blockchain&\footnotesize PRE&\footnotesize BC, FL&\3&\1&\1\\
\footnotesize\cite{[88]}&\footnotesize Combined blockchain&\footnotesize PRE&\footnotesize BC, Signcryption&\3&\2&\1\\
\footnotesize\cite{[90]}&\footnotesize Combined blockchain&\footnotesize PRE&\footnotesize CC, BC&\3&\1&\1\\
\footnotesize\cite{[313]}&\footnotesize Combined blockchain&\footnotesize PRE&\footnotesize BC&\3&\2&\1\\
\footnotesize \cite{[31]} & \footnotesize Cloud computing & \footnotesize PRE & \footnotesize CC & \2 & \2 & \2 \\
\footnotesize\cite{[56]}&\footnotesize Cloud computing&\footnotesize PRE, ABE&\footnotesize CC, ZKP&\3&\1&\1\\
\footnotesize \cite{[84]} & \footnotesize Cloud computing & \footnotesize PRE & \footnotesize CC, IBE & \3 & \1 & \3 \\
\footnotesize \cite{[95]} & \footnotesize Cloud computing & \footnotesize PRE & \footnotesize CC, BE & \2 & \2 & \1 \\
\footnotesize\cite{[55]}&\footnotesize Combined with cryptography knowledge&\footnotesize PRE&\footnotesize -&\2&\2&\1\\
\footnotesize\cite{[87]}&\footnotesize Combined with cryptography knowledge&\footnotesize PRE&\footnotesize IBE&\3&\1&\1\\
\footnotesize\cite{[21]}&\footnotesize Combined with cryptography knowledge&\footnotesize PRE, ABE&\footnotesize CC&\2&\1&\2\\
\footnotesize\cite{[47]}&\footnotesize Cryptography knowledge and HE&\footnotesize PRE&\footnotesize HE&\3&\1&\3\\
\footnotesize\cite{[66]}&\footnotesize Combined with cryptography knowledge&\footnotesize PRE, SE&\footnotesize BC, PQC&\3&\1&\1\\
\footnotesize\cite{[101]}&\footnotesize Combined with cryptography knowledge&\footnotesize PRE, ABE&\footnotesize CC&\3&\1&\1\\
\footnotesize\cite{[107]}&\footnotesize Combined with cryptography knowledge&\footnotesize PRE&\footnotesize CC&\3&\1&\1\\
\footnotesize\cite{[99]}&\footnotesize Combined with cryptography knowledge&\footnotesize PRE&\footnotesize CC&\3&\1&\2\\
\footnotesize\cite{[6]}&\footnotesize Combined with HE&\footnotesize PRE &\footnotesize HE&\3&\1&\1\\
\footnotesize\cite{[16]}&\footnotesize Combined with HE&\footnotesize PRE&\footnotesize CC, IBE, PQC&\3&\1&\1\\

\end{longtable}
\renewcommand{\arraystretch}{1.0}
\subsection{Security Improvement Based on ABE}
ABE achieves fine-grained access control by linking data access rights to user attributes, allowing only users with specific attributes to decrypt and access data. 
In recent years, there has been a continuous emergence of ABE schemes incorporating blockchain, multi-authority, hidden strategies, and anonymity. These schemes not only enhance the system's anti-attack capabilities but also adapt to dynamic requirements across different scenarios, offering a safer and more reliable solution for data sharing.
\subsubsection{ABE Based on Distributed or Multi-Attribute Authorities}
%Blockchain provides enhanced security, fine-grained access control and the ability of automatic execution for data sharing based on attribute encryption through its decentralized, tamper-proof and intelligent contract characteristics, thus realizing the safe sharing and efficient management of data.\cite{[5],[9],[14],[42]} Dynamically manage data access rights through smart contracts to ensure the security and flexibility of data sharing. Velmurugan et al.\cite{[33]} combine attribute-based encryption and searchable encryption technology to realize fine-grained access control and keyword search for encrypted medical data. Through blockchain technology, the master key is decentralized and managed on all attribute nodes to ensure that the system can still operate normally even if some nodes are breached. Xu et al\cite{[72]} constructed a new medical data sharing scheme based on blockchain, which broke the system boundary and realized cross-hospital diagnosis. Based on attribute encryption, an authorization and revocation mechanism was designed, which allowed hospitals to authorize or revoke doctors' access rights to data, ensuring flexible data access control, while using zero-knowledge proof protocol to ensure the credibility of hospital matching algorithm and the privacy protection of patients for medical reports. Dai et al.\cite{[80]} embedded the zero-knowledge proof protocol into the blockchain, and used the decentralized characteristics of the blockchain to avoid the forgery of centralized verification.
BC offers ABE decentralized, tamper proof features and smart contract based automatic execution, enhancing data sharing security and enabling efficient management. References \cite{[9],[42]} realize the dynamic data access control via smart contracts, ensuring data sharing security and flexibility. Xu et al. \cite{[72]} introduced an authorization mechanism that combines BC and ABE. The scheme incorporates a many-to-many matching mechanism. Through this mechanism, patients' health data can be represented by multiple keywords, while doctors' areas of expertise can be represented by multiple interests. Dai et al. \cite{[80]} incorporated ZKP into BC, using its decentralized nature to prevent centralized verification forgery.\par
%%Velmurugan et al. \cite{[33]} integrated ABE and SE to achieve fine grained access control and keyword search for encrypted medical data. By leveraging BC technology to decentralize the master key across all attribute nodes, they ensured normal system operation even if some nodes were compromised. 
%\begin{figure*}[!t]
%    \centering
%    \includegraphics[width=12cm]{p4.png}
%    \caption{The data flow of the privacy-preserving medical data sharing scheme}
%    \label{p4}
%\end{figure*}

In ABE, introducing multi-authority institutions aims to address the challenges of single-authority schemes, particularly collusion prevention. In multi-authority ABE \cite{[93]}, multiple authorities work together in key generation and distribution. Each authority manages a unique set of attributes. This collaborative approach not only enhances the system's flexibility and scalability but also strengthens its security and privacy protection capabilities. Wang et al. \cite{[4]} utilized multiple attribute authorization bodies to manage attributes. This design avoids single points of failure, enables patients to create custom access policies, and gives them greater control over their personal data. Duan et al. \cite{[7]} incorporated multi-authority ABE technology, allowing several independent attribute authorities to manage attributes and distribute keys. When combined with PRE, this approach introduces agents. This enables DO to delegate access rights without disclosing sensitive information. Velmurugan et al. \cite{[33]} employed data fingerprinting technology. This technology generates short fingerprints for shared data. It enables data checks before cloud storage, thereby helping to prevent data leakage.

\subsubsection{ABE Based on Hiding Strategy}
%The tamper-proof feature of blockchain ensures the integrity and authenticity of data and keeps the data valuable in the sharing process. However, the transparency of blockchain may lead to the disclosure of access structure and user attributes. Attribute encryption with hidden policies can ensure fine-grained access control in the data sharing process and protect user privacy by hiding access policies and attributes. Yang et al.\cite{[82]} hide the attributes in the access policy by confusing the attribute mapping function, which effectively prevents the specific attribute values of users from being leaked to the third party and ensures the privacy of users. Yang et al.\cite{[14]} hide the attributes in the access policy by transforming the access policy into a vector-matrix form, and prevent the disclosure of private information of authorized users. Yang et al.\cite{[45]} protected users' privacy by polynomial and vector representation, and ensured that the user's attribute set met the access policy by a verification method combining bilinear pairing and predicate encryption, while hiding the policy and attribute values. Cai et al.\cite{[50]} hide the access policy of the data owner and protect privacy by embedding the access policy with random numbers.
BC's tamper proof nature ensures data integrity and authenticity, preserving data value during sharing. However, its transparency may expose access structures and user attributes. ABE with hidden policy can achieve fine-grained access control while protecting user privacy by concealing access policies and attributes. Yang et al. \cite{[82]} concealed attributes in the access policy by obfuscating the attribute mapping function. This prevents user-specific attribute values from being disclosed to third parties, safeguarding user privacy. Yang et al. \cite{[14]} transformed the access policy into a vector-matrix format to mask attributes, preventing the leakage of authorized users' private information. Yang et al. \cite{[45]} employed polynomials and vector representation to shield user privacy. They verified whether a user's attribute set satisfied the access policy using a method that combines Bilinear Pairing with Predicate Encryption, all while keeping the policy and attribute values hidden. Cai et al. \cite{[50]} embedded the DO's access policy with random numbers to hide it and protect privacy. Wu et al. \cite{[49]} introduced inner product predicate to realize completely hidden access strategy and prevent sensitive medical data from leaking. POOMEKUM et al. \cite{[323]} embedded the salted attribute-hashing Mechanism into the linear secret sharing scheme matrix can hide the real access policy even in the process of fog node processing.
\subsubsection{Anonymous ABE}
%Anonymous ABE is an attribute-based encryption scheme, which not only protects the confidentiality of data, but also ensures the anonymity of the receiver. In anonymous ABE, the user's attributes are hidden, which makes it impossible for attackers to infer the user's attribute information through ciphertext or key. Zhang et al.\cite{[23]} realized the double anonymity of data owner and data user by combining anonymous verifiable algorithm and attribute encryption algorithm, and protected the identity privacy of both parties. Children's books support the accountability of malicious users and can revoke their rights, which enhances the trust of data sharing environment. Zhang et al.\cite{[44]} improved the anonymous CP-ABSE\cite{[7]}, which can realize access policy hiding and flexible access control of encrypted data, and at the same time make the receiver anonymous. Chen et al.\cite{[100]} preprocessed the medical data by K- anonymity technology, blurred the quasi-identifier (QID) that may expose the patient's identity, and ensured the anonymity of the data. Yang et al. [110] allowed the signer to use a set of attributes instead of identity through the attribute-based signature protocol, which not only verified the authenticity of the data source, but also protected the identity privacy of the signer.
In anonymous ABE, user attributes are concealed, preventing attackers from inferring attribute information from ciphertext or keys. Zhang et al. \cite{[23]} integrated an anonymous verifiable algorithm with ABE to achieve double anonymity for users, protecting both parties' identity privacy. Their approach also supports accountability for malicious users and enables right revocation, bolstering data sharing environment trust. Zhang et al. \cite{[44]} enhanced anonymous CP-ABSE by hiding access policies and enabling flexible encrypted data control, while maintaining receiver anonymity. Chen et al. \cite{[100]} preprocessed medical data with K-anonymity technology. They blurred quasi-identifiers that might reveal patient identities, ensuring data anonymity. 
%Yang et al. \cite{[110]} employed an attribute based signature protocol, allowing signers to use attribute sets instead of identities. This method verified data source authenticity while protecting signer identity privacy.
\subsection{Security Improvement Based on SE}
%In recent years, the rapid development of enhancement technology has opened up a new direction for realizing efficient and secure search, dynamic data update and complex query management in multi-user environment.
Traditional SE-based data sharing schemes encounter numerous challenges concerning security, privacy, and efficiency. Recently, the rapid advancement of BC, privacy computing technologies, and cryptographic foundations has offered novel approaches to enhance data sharing schemes security. Integrating these cutting-edge technologies has enabled more efficient and secure search functions in multi-user settings, as well as dynamic data updates and complex query management, thereby paving a new way for secure data sharing.
\subsubsection{SE Combined with Blockchain Technology}Traditional SE has several practical problems like fragile key management, centralized data storage risks, and search process privacy leakage. BC technology offers new solutions for SE's data security and privacy protection. 
Shamshad et al. \cite{[102]} proposed medical data sharing schemes that combine SE with BC. This improves data sharing efficiency and security while protecting patient privacy. Huang et al. \cite{[18]} created a blockchain assisted cloudy storage model. It enables industrial data indirect retrieval via alliance chains, reducing single  point failure risks and enhancing industrial data storage safety. Zhou et al. \cite{[70]} presented a blockchain based lightweight SE sharing scheme for in vehicle social networks. It ensures efficient and secure data matching and sharing. Lu et al. \cite{[25]} proposed a new searchable PRE scheme, which allows data buyers to verify the validity of search results without exposing the decryption key of the DO. And based on the smart contract of blockchain, the automation and fairness of data transactions in the Internet of Things are realized.
%Yang et al. \cite{[78]} utilized blockchain's dynamic consensus committee to generate temporary partial keys, avoiding single point  failure and single key exposure problems. Jihyeon Oh et al. \cite{[81]} applied blockchain to fog nodes. This ensures keyword verification task security and data transparency during searches, and promotes safe Internet of Things(IoT) environment data sharing. Xu et al. \cite{[115]} used blockchain as a retrieval platform and proposed a blockchain based dynamic searchable symmetric encryption scheme. It addresses data sharing and security issues in multi-user scenarios. 

\subsubsection{SE Combined with Privacy Computing Technology}The integration of SE with privacy computing technologies offers a more reliable solution for data sharing and retrieval in complex environments. Muazu et al. \cite{[162]} employed MPC technology to ensure that participants cannot access each other's raw data during data fusion, safeguarding gradient parameters in FL systems. Building on MPC, Wang et al. \cite{[79]} used Secure Multiparty Computation and Pseudorandom Function to generate search indexes, shared tokens, and search tokens. This approach supported dynamic user management, including additions and revocations, thereby ensuring the security and real-time performance of data sharing. Beyond MPC, the incorporation of DP technology into SE allows for the introduction of controllable noise during queries. This effectively hides keywords and access patterns, thereby enhancing security. Saidi et al. \cite{[123]} significantly improved model accuracy while protecting privacy through DP technology and designed a personalized DP mechanism. This mechanism allowed clients to add Laplacian noise to model gradients based on a customized privacy budget. Chen et al. \cite{[15]}, leveraging DP's strong security, combined DP with SE. They protected patient identity privacy by adding local noise to data, differentiating between identity and data privacy. They preprocessed patient identity data using DP technology and uploaded perturbed data to cloud servers to guard against attacks by untrusted third parties. When combined with ZKP technology, SE can verify the legality of search requests and data access without revealing additional information. 
%Zeng et al. \cite{[144]} innovatively merged ZKP with key agreement technology within a blockchain assisted data governance framework. They designed the BCDS-ZK scheme to verify data ownership while ensuring confidentiality and anonymity. This facilitates efficient and secure cross domain data sharing.
\subsubsection{SE Combined with Cryptography Knowledge}
By integrating cryptographic principles, SE can protect data privacy through encryption algorithms and support secure searches on encrypted data. Zhou et al. \cite{[2]} introduced the Cryptographic Reverse Firewall (CRF) technology. Deployed in trusted areas, CRF randomizes users' keys, indexes, and search trapdoors, boosting resistance to backdoor attacks and protecting the security of these elements. Yin et al. \cite{[38]} proposed a new three step interactive key generation algorithm. In it, user private key fragments are generated through interaction with all attribute nodes. This process conceals user identity information from attribute nodes, ensuring the privacy of both the information and the key fragments. Lattice based cryptography, especially schemes based on the Learning With Errors (LWE) hypothesis (\cite{[59],[117]}), excels in resisting quantum computer attacks. 
Zhang et al. \cite{[106]} effectively prevented the cloud server from obtaining data privacy through Internal Keyword Guessing Attack (IKGA) by introducing authorization mechanism and technology of resisting IKGA.

 %Liu et al. \cite{[46]} enhanced intermediate ciphertext encryption security via the Efficient and Secure Key Issuing Identity-Based Encryption (ESKI-IBE) protocol, effectively preventing sensitive medical data leakage and ensuring medical data privacy and security. 
\subsection{Security Improvement Based on PRE}
The integration of BC, CC, cryptographic foundation, and HE technology with PRE enhances PRE's security in data sharing and drives its adoption in more intricate scenarios.
\subsubsection{PRE Combined with Blockchain Technology}
The integration of PRE with smart contracts enables secure data storage and access authentication, effectively safeguarding data privacy and security. In the realm of privacy protection, Huang et al. \cite{[103]} realized the safe sharing of patients' medical data under privacy protection by combining ZKP and BC technology, and ensured the availability, consistency and traceability of the data. Prem et al. \cite{[111]} put forward a secure electronic health cloud framework, ensuring data privacy protection alongside efficient data sharing. Turning to the area of data security, Mittal et al. \cite{[62]} designed a novel group key management mechanism that utilizes blockchain to secure group encryption, thereby protecting the internal communications of medical teams and preventing data leakage and unauthorized access.Ahene et al.\cite{[88]} use the distributed account book characteristics of blockchain to record all the access and modification operations of Electronic Health Record (EHR) data, provide transparent audit logs, and enhance the auditability of the system. Lastly, Sammeta et al. \cite{[90]} presented an EHR sharing scheme based on blockchain and chaotic re-encryption. By harnessing the decentralized nature of blockchain and the high security of chaotic encryption, this scheme ensures the integrity and security of data. Garcia et al. \cite{[313]} put forward a decentralized data governance framework that integrates BC, PRE and BBS signatures, so that DO can control data access rights and accurately share specific data attributes with the help of selective disclosure.
%Hossen et al. \cite{[60]} introduced a verifiable local model aggregation method that integrates an improved Certificate-Based Proxy Re-Encryption with BC and FL mechanisms. By refining the CB-PRE algorithm, they bolstered key management security, facilitating data sharing and model training among multiple medical institutions without compromising the privacy of local data. 
%Chen et al. \cite{[47]} developed a privacy preserving data aggregation scheme based on  Ring Learning With Errors (R-LWE). This approach not only ensures post-quantum security but also achieves robust privacy protection, fault tolerance, and verifiable aggregation functions. notable works include that of Guo et al. \cite{[24]}, who proposed a decentralized authentication mechanism leveraging smart contracts to shield user privacy and eliminate dependence on third-party centralized authentication services. Wang et al. \cite{[34]} proposed the hybrid blockchain health data sharing (HSHB) scheme, which merges private and consortium blockchains. By categorizing sharing transactions according to different sharing entities, this scheme averts interference between entities and effectively addresses security challenges in health data sharing. Raghav et al. \cite{[67]} combined the strengths of public and private blockchains to provide decentralized identity authentication, data integrity, and non-tamperability, supporting large-scale data sharing.
\subsubsection{PRE Combined with Cloud Computing}
CC ensures data confidentiality, integrity, and fine-grained access control, yet still faces challenges such as data security, user privacy protection, and key revocation. To address these, Reddy et al. \cite{[31]} combined PRE with ABE and introduced a trust and reputation model. This approach significantly enhances data sharing security and effectively prevents data leakage, tampering, and illegal access. Zhao et al. \cite{[56]} combined the ABE and PRE technologies for EHRs in E-health Cloud, which realized the dynamic fine-grained access control of encrypted electronic health records, supported the big universe attribute space, partial hiding strategy and verifiability, effectively protected the privacy of patients and improved the security of the system. Deng et al. \cite{[84]} propose the policy-based broadcast access authorization scheme, which integrates identity-based broadcast encryption and key-policy attribute-based encryption into PRE to address the flexibility issue in multi-user data sharing in clouds, achieving secure and efficient data sharing. Ge et al. \cite{[95]} introduce a new key generation and revocation algorithm, enabling proxy servers to revoke designated receivers' keys without exposing DO's private keys.
%It resolves key revocation issues in existing Identity-Based Broadcast Proxy Re-Encryption schemes and addresses identity based encrypted data sharing challenges in cloud environments regarding flexibility, expansibility, and key revocation.
%Shaik et al. \cite{[6]} built an hybrid PRE (HPRE) scheme on the learning with linear equations errors assumption. Integrating HE and PRE benefits, it allows encrypted data computation and flexible sharing among different users, further strengthening information integrity and confidentiality. 
%Li et al. \cite{[16]} proposed a lattice-based Ciphertext-Linear Proxy Re-Encryption (CL-PRE) scheme that resists quantum computing attacks and effectively guarantees big data safe sharing in cloud computing environments. Reddy et al.
%Jaya et al. \cite{[116]} proposed a comprehensive scheme integrating IBE, PRE, Information-Centric Networking (ICN), and blockchain technology. This multi-dimensional approach ensures data confidentiality, integrity, and privacy protection, greatly improves cloud computing security in the PRE field, and offers innovative solutions for enhancing data sharing security in cloud environments.
\subsubsection{PRE Combined with Cryptography Knowledge}
Group cryptography holds significant advantages in multi-user collaboration and group communication, effectively meeting complex network environment secure communication needs. Xia et al. \cite{[55]} achieved high efficiency and security in resource-limited environment by using elliptic curve cryptography and optimized PRE scheme. Ren et al. \cite{[87]} combined certificateless encryption with autonomous path PRE. This approach bypasses traditional certificate management complexities, allows data owners to fully control access paths. Mittal et al. \cite{[62]} introduced a blockchain based group encryption mechanism to ensure secure medical team data sharing and prevent leakage risks. Sun et al. \cite{[21]} applied key blinding technology to shield ABE users' private keys. This prevents cloud server and Public Key Infrastructure (PKI) user collusion attacks. 
Nonetheless, group cryptography is vulnerable to quantum attacks. To address this, Wu et al. \cite{[47]} employed ideal lattices to instantiate homomorphic threshold PRE. This ensures post-quantum security while maintaining lightweight computational complexity. 
%Shaik et al. \cite{[6]} improved and extended the Brakersky-Gentry-Vaikuntanathan (BVG) algorithm. Their version retained the original's ability to evaluate L-level arithmetic circuits and perform boot operations while boosting the scheme's overall security and efficiency. Li et al. \cite{[16]} utilized lattice based problems like LWE to construct an encryption scheme, ensuring quantum computing attack resistance. 
Zhu et al. \cite{[66]} achieved secure quantum key forwarding. By integrating the inherent true randomness and high security of quantum keys with PRE technology, their approach effectively supports secure multi-user sharing.
Ge et al. \cite{[101]} pioneered the use of bilinear pairing in constructing efficient verifiable and fair attribute-based PRE schemes. Their work ensures that users can verify the correctness of re-encrypted ciphertexts while shielding cloud servers from malicious accusations.
Li et al. \cite{[107]} proposed a scheme combining PRE and public key encryption with equality test, which can realize data search and secure sharing across different public key encryption on the premise of protecting privacy. Guo et al. \cite{[99]} designed a non-interactive APRE scheme. By embedding the agent's private key information in the re-encryption key, the re-encryption key is bound to the agent's key pair, thus realizing the traceability of the agent's behavior.
\subsubsection{PRE Combined with HE Technology}
HE enables specific computations on encrypted data without decryption, while PRE allows for flexible access rights transfer of encrypted data between users. By organically combining these two technologies, a more powerful data processing and sharing mechanism can be constructed. Shaik et al. \cite{[6]} proposed an homomorphic PRE scheme based on the LWE assumption, integrating the benefits of HE and PRE. Their scheme supports computations on encrypted data and enables flexible sharing of encrypted data among different users. Li et al. \cite{[16]} extended the certificateless public key encryption scheme to the certificateless PRE scheme, which supports additive homomorphism for both original and re-encrypted ciphertexts, thus greatly improving data processing flexibility and efficiency. Wu et al. \cite{[47]} presented a homomorphic threshold PRE scheme with re-encryption verifiability. It achieves strong privacy protection, fault tolerance, and verifiable aggregation with only linear operations involved. After ideal lattice instantiation, the scheme can also resist quantum attacks. Prem et al. \cite{[111]} proposed a secure electronic health cloud framework combining HE and PRE, employing the Boneh-Goh-Nissim cryptosystem to realize HE functionality and ensure secure cloud data storage.   

%\subsection{Discussion}
%The lack of anti-quantum capability is the core underlying security flaw in current data sharing schemes based on SE, PRE, and ABE, rooted in their reliance on classical mathematical problems such as discrete logarithm (DL) and bilinear pairing (BP). These problems can be efficiently cracked by Shor and Grover algorithms once quantum computing matures. In terms of technical characteristics, the trapdoor generation and index matching of SE, re-encryption key derivation and ciphertext conversion of PRE, and attribute trapdoor construction and decryption verification of ABE all depend on the above mathematical primitives to achieve "a balance between security and functionality", which further exacerbates quantum security risks.\par
%To address this issue, future efforts will focus on lattice-based cryptography optimization: the SE scheme can construct a trapdoor and index mechanism based on RLWE, leveraging lattice-based quantum resistance to ensure security; the PRE scheme adopts ideal lattices or NTRU lattices to design anti-quantum re-encryption keys, guaranteeing the confidentiality of keys and ciphertexts; the ABE scheme needs to break through lattice-based attribute trapdoor construction technology, utilizing lattice-based homomorphic properties to match policies and attributes, while resisting Grover algorithm attacks to ensure verification security.

\section{Performance Improvement of Data Sharing}\label{s5}
On the premise of ensuring data security and privacy, ABE, SE and PRE reduce the computational and storage overhead through optimized storage structures, retrieval efficiency, and streamlined workflows, and greatly improve the overall performance of data sharing. This section systematically reviews and analyzes existing research on how ABE, SE, and PRE enhance data sharing performance, as detailed in Table \ref{table3}. 
%\Lv{Reorder according to the order of classification and appearance}
\renewcommand{\arraystretch}{1.0}
\subsection{Performance Improvement Based on ABE}
In order to optimize the data sharing performance of ABE, the current research mainly focuses on designing outsourcing ABE, lightweight ABE and optimizing access policy architecture. These schemes significantly improve the efficiency of data sharing by improving the encryption algorithm and optimizing the calculation task allocation process, so that it can better adapt to complex and changeable application scenarios.

\subsubsection{ABE Based on Outsourcing}
%%Outsourcing attribute encryption is to outsource some or all of the decryption tasks to a trusted third party by using outsourcing decryption mechanism, thus reducing the local computing pressure, especially for devices with limited computing resources. In the multi-user and dynamic permission scenario, it can flexibly adjust the user's permissions, without updating and distributing a large number of keys, and it is easy to expand the user group and service scope. Some existing studies\cite{[17],[14],[110]}  outsource complex operations to cloud service providers by improving the data sharing framework to achieve lightweight decryption, thus significantly reducing the computational burden of users. Miao et al. \cite{[22]} split the secret key and distributed it to the cloud server, which enabled the cloud server to perform most time-consuming encryption operations, and introduced a short signature mechanism to verify whether the cloud server correctly performed encryption operations. Zhao et al.\cite{[27]} proposed a completely outsourced and completely verifiable scheme that supports the generation, encryption and decryption of verifiable outsourced keys, without multiple rounds of interaction or high verification cost.Yang et al.\cite{[82]}outsourced the complex decryption calculation to a third-party cloud server, and stored the verification parameters by using the tamper-proof characteristics of the blockchain, which can be used by data users to verify the correctness of the third-party outsourced decryption results.
Outsourced ABE leverages third party decryption to alleviate local computational burdens, particularly benefiting devices with limited resources. In multi-user and dynamic permission settings, it adjusts user permissions with minimal key updates. References \cite{[17],[37]} enhance data sharing frameworks by outsourcing complex operations to cloud providers, lightening user computational loads. Miao et al. \cite{[22]} split secret keys and distributed them to cloud servers, enabling these servers to handle most time consuming encryption tasks. They also introduced short signatures to verify server side encryption operations. Zhao et al. \cite{[27]} proposed a fully outsourced and verifiable scheme supporting outsourced key generation, encryption, and decryption, without multi-round interactions or high verification costs. Yang et al. \cite{[82]} outsourced complex decryption to third-party cloud servers and used blockchain's tamper-proof nature to store verification parameters. This allows data users to verify third-party decryption results.

\renewcommand{\arraystretch}{0.1}
\begin{longtable}{
>{\raggedright\arraybackslash}p{0.6cm} 
>{\raggedright\arraybackslash}p{4.6cm} 
>{\raggedright\arraybackslash}p{1.6cm} 
>{\raggedright\arraybackslash}p{2cm}
>{\centering\arraybackslash}p{0.7cm} 
>{\centering\arraybackslash}p{0.9cm}
>{\centering\arraybackslash}p{0.9cm}}
\caption{Summary of Research on Improving the Performance of Data Sharing by ABE, SE and PRE}
\label{table3}  \\
\hline
\footnotesize \textbf{Paper} & 
\footnotesize \textbf{Research Contents} & 
\footnotesize \textbf{Encryption Technology} & 
\footnotesize \textbf{Enhancement Technology}& 
\footnotesize \textbf{Secu\-rity} & 
\footnotesize \textbf{Perfor\-mance} & 
\footnotesize \textbf{Functi\-onality}  
\\

\hline
\endhead 
\addlinespace[2pt] %
\hline
\endfoot 
\hline
\addlinespace[2pt] %
\multicolumn{7}{l}{\footnotesize Note: The comparison of security strength,  performance and functionality is expressed as follows: \3 > \2 > \1.} 
%, indicating that variable \3 has the strongest influence on the dependent variable

\endlastfoot  
\addlinespace[2pt] %
\footnotesize \cite{[17]} & \footnotesize Outsourcing ABE, lightweight ABE & \footnotesize ABE & \footnotesize CC, BC, Sig & \1 & \2 & \2 \\
\footnotesize \cite{[37]} & \footnotesize Outsourcing ABE & \footnotesize ABE & \footnotesize BC & \2 & \3 & \1 \\
\footnotesize \cite{[22]} & \footnotesize Outsourcing ABE & \footnotesize ABE & \footnotesize CC & \1 & \3 & \1 \\
\footnotesize \cite{[27]} & \footnotesize Outsourcing ABE & \footnotesize ABE & \footnotesize CC & \1 & \3 & \2 \\
\footnotesize \cite{[82]} & \footnotesize Outsourcing ABE & \footnotesize ABE & \footnotesize BC & \2 & \2 & \2 \\
\footnotesize \cite{[68]} & \footnotesize Lightweight ABE & \footnotesize ABE & \footnotesize CC & \1 & \2 & \2 \\
\footnotesize \cite{[92]} & \footnotesize Lightweight ABE & \footnotesize ABE & \footnotesize Signcryption, CC & \1 & \3 & \3 \\
\footnotesize \cite{[97]} & \footnotesize Lightweight ABE & \footnotesize ABE, SE & \footnotesize CC & \1 & \3 & \1 \\
\footnotesize \cite{[20]} & \footnotesize Lightweight ABE & \footnotesize ABE, SE & \footnotesize BC & \2 & \3 & \1 \\
\footnotesize \cite{[43]} & \footnotesize Lightweight ABE & \footnotesize ABE, SE & \footnotesize CC & \2 & \3 & \1 \\
\footnotesize\cite{[13]}&\footnotesize Optimize the access policy structure &\footnotesize ABE&\footnotesize CC&\2&\2&\1\\
\footnotesize \cite{[64]} & \footnotesize Optimize the access policy structure & \footnotesize ABE, PRE & \footnotesize CC, IBE & \1 & \2 & \2 \\
\footnotesize \cite{[105]} & \footnotesize Optimize the access policy structure & \footnotesize ABE & \footnotesize CC & \1 & \3 & \1 \\
\footnotesize\cite{[5]}&\footnotesize Optimize the access policy structure &\footnotesize ABE&\footnotesize BC&\2&\2&\1\\
\footnotesize \cite{[8]} & \footnotesize Optimize the access policy structure & \footnotesize ABE & \footnotesize CC & \1 & \3 & \1 \\
%\hline
\footnotesize \cite{[15]} & \footnotesize Storage optimization & \footnotesize SE & \footnotesize BC, CC & \2 & \2 & \2 \\
\footnotesize\cite{[12]}&\footnotesize Storage optimization &\footnotesize SE, PRE&\footnotesize BC, CC &\2&\2&\1\\
\footnotesize \cite{[52]} & \footnotesize Storage optimization & \footnotesize SE, ABE, PRE & \footnotesize BC, IPFS & \1 & \3 & \2 \\
\footnotesize \cite{[61]} & \footnotesize Storage optimization & \footnotesize SE, PRE & \footnotesize IPFS, BC & \1 & \3 & \1 \\
\footnotesize \cite{[94]} & \footnotesize Storage optimization & \footnotesize SE, ABE  & \footnotesize IPFS, BC & \1 & \3 & \1 \\
\footnotesize \cite{[1]} & \footnotesize Query optimization & \footnotesize SE & \footnotesize HE & \2 & \3 & \1 \\
\footnotesize \cite{[96]} & \footnotesize Query optimization & \footnotesize SE & \footnotesize - & \1 & \2 & \2 \\
\footnotesize \cite{[3]} & \footnotesize Query optimization & \footnotesize SE, ABE & \footnotesize BC & \2 & \3 & \1 \\
\footnotesize \cite{[26]} & \footnotesize Query optimization & \footnotesize SE, PRE & \footnotesize CC & \2 & \3 & \1 \\
\footnotesize\cite{[115]} & \footnotesize Lightweight SSE & \footnotesize SE & \footnotesize BC & \2 & \3 & \3 \\
\footnotesize\cite{[81]} & \footnotesize Lightweight SSE & \footnotesize SE & \footnotesize CC & \2 & \3 & \3 \\
\footnotesize\cite{[76]} & \footnotesize Lightweight SSE & \footnotesize SE & \footnotesize CC & \2 & \3 & \3 \\
\footnotesize \cite{[78]} & \footnotesize Lightweight SSE & \footnotesize SE & \footnotesize CC, BC & \2 & \3 & \1 \\
%\footnotesize \cite{[109]} & \footnotesize Lightweight SSE & \footnotesize SE & \footnotesize CC & \1 & \3 & \1 \\
%\hline
\footnotesize \cite{[29]} & \footnotesize Binding IBE & \footnotesize PRE & \footnotesize IBE, BC & \2 & \3 & \1 \\
\footnotesize \cite{[73]} & \footnotesize Binding IBE & \footnotesize PRE & \footnotesize IBE, CC & \1 & \2 & \2 \\
\footnotesize \cite{[86]} & \footnotesize Binding IBE & \footnotesize PRE & \footnotesize IBE & \1 & \2 & \1 \\
\footnotesize \cite{[11]} & \footnotesize Binding IBE & \footnotesize PRE, ABE & \footnotesize IBE, CC & \1 & \2 & \2 \\
\footnotesize \cite{[46]} & \footnotesize Binding IBE & \footnotesize PRE & \footnotesize IBE, BC, Sig & \2 & \3 & \--\\
\footnotesize \cite{[322]} & \footnotesize Binding IBE & \footnotesize PRE & \footnotesize IBE, PQC & \2 & \3 & \--\\
\footnotesize\cite{[54]} & \footnotesize Binding BE & \footnotesize PRE & \footnotesize BC, BE & \1 & \2 & \2 \\
\footnotesize \cite{[201]} & \footnotesize Binding BE & \footnotesize PRE & \footnotesize BE, IBE & \2 & \3 & \1 \\
\footnotesize \cite{[95]} & \footnotesize Binding BE & \footnotesize PRE & \footnotesize CC, BE & \2 & \2 & \1 \\
\footnotesize \cite{[31]} & \footnotesize Lightweight PRE & \footnotesize PRE & \footnotesize CC & \2 & \2 & \2 \\
\footnotesize\cite{[67]}&\footnotesize Lightweight PRE &\footnotesize PRE &\footnotesize BC &\2&\2&\2\\
\footnotesize \cite{[32]} & \footnotesize Lightweight PRE, trusted third party & \footnotesize PRE & \footnotesize BC & \2 & \3 & \2 \\
\footnotesize \cite{[71]} & \footnotesize Lightweight PRE & \footnotesize PRE & \footnotesize BC & \2 & \3 & \1 \\
\footnotesize \cite{[36]} & \footnotesize Lightweight PRE & \footnotesize PRE & \footnotesize BC, IPFS & \1 & \3 & \1 \\
\footnotesize \cite{[41]} & \footnotesize Lightweight PRE & \footnotesize PRE, ABE & \footnotesize - & \2 & \3 & \1 \\
\footnotesize \cite{[24]} & \footnotesize Trusted third party & \footnotesize PRE & \footnotesize IPFS, BC, ZKP & \2 & \2 & \2 \\
\footnotesize \cite{[34]} & \footnotesize Trusted third party & \footnotesize PRE & \footnotesize BC & \2 & \3 & \1 \\

\end{longtable}

\subsubsection{Lightweight ABE}
Lightweight ABE reduces computational resource usage by simplifying encryption and decryption processes \cite{[17],[23]} or optimizing access policy structures \cite{[50],[68]}, enhancing data sharing efficiency and access control. Yao et al. \cite{[17]} employed lightweight symmetric encryption algorithms to secure data transmission between terminal devices and edge nodes. By directly encrypting raw data based on attributes, they considerably boosted system computation efficiency. Cui et al. \cite{[68]} assigned users with identical attributes to the same group and linked them to corresponding proxy servers, minimizing user interaction and proxy server conversion costs. Obiri et al. \cite{[92]} enhanced system performance by optimizing the attribute-based signcryption scheme, reducing ciphertext size and computational costs. Bao et al. \cite{[97]} developed a lightweight ABSE scheme, significantly reducing the computational load on resource constrained patients and doctors. Xie et al. \cite{[20]} designed an Internet of Things (IoT) suitable lightweight data sharing scheme. By integrating inverted index structures and smart contract technology, it enables efficient data sharing on resource limited IoT devices. Wang et al. \cite{[43]} proposed a lightweight key aggregation authorization SE scheme. Effectively countering ``key forgery'' and ``trapdoor forgery'' attacks, it ensures data security and privacy while enabling efficient authorization and keyword searching.
 %Zhang et al. \cite{[23]} replaced computation intensive bilinear pairing operations with elliptic curve operations, reducing terminal equipment calculation and communication burdens. They also designed a batch data integrity verification algorithm to improve verification efficiency. Cai et al. \cite{[50]} introduced a CRT-based dynamic multi-group management scheme. It supports group merging and splitting while keeping users' private keys unchanged, thus lowering group administrator computational overhead. 
\subsubsection{Optimize the Structure of Access Policies}
Many researchers have enhanced ABE's data sharing performance by innovating access policy structures. Yang et al. \cite{[13]} integrated multiple access policies into a unified one via a hierarchical access structure, reducing encryption and decryption computational overhead. Wang et al. \cite{[64]} made system initialization algorithms and common parameter sizes independent of attribute numbers using the bilinear entropy extension lemma. This kept computational costs stable during encryption, decryption, and re-encryption even as attributes increased. Zhi et al. \cite{[105]} applied the EM algorithm to cluster IoT device data, reducing redundancy and boosting processing efficiency, and then used ABE to ensure data security. Chen et al. \cite{[5]} proposed a novel multi-attribute sketch storage method. Storing encrypted medical data hash values in blockchains, it cuts storage costs and improves query speed. Thushara et al. \cite{[8]} introduced a session based data sharing mechanism. By creating sessions between fog nodes, it enhances data transmission efficiency and reduces data sharing delays and storage burdens. 

\subsection{Performance Improvement Based on SE}
As data scales expand and application requirements grow, SE optimizes storage, enhances query efficiency, and cuts computation and communication costs via storage structure improvements, query optimizations, and lightweight SSE technology. These advancements significantly boost SE's performance in large-scale data processing and broaden its application scope.
\subsubsection{Storage Optimization}
SE faces performance bottlenecks when dealing with large-scale data storage and retrieval. Storage optimization measures, such as adopting block storage and optimizing index structures, can effectively reduce storage overhead and improve retrieval efficiency, thereby significantly enhancing SE's performance. Chen et al. \cite{[15]} describe a chain-on-chain and chain-off storage method, which stores security indexes via private and alliance chains respectively to prevent collusion attacks. Feng et al. \cite{[12]} combined blockchain with private cloud storage, allowing industrial enterprises to store large amounts of sensitive data in private clouds. Classified metadata is stored in different blockchains, ensuring data security and reducing the storage burden on the blockchain. References \cite{[52]}, \cite{[61]}, and \cite{[94]} indicate that the original data ciphertext is stored in IPFS. This approach ensures data security and efficient access, solves the single point failure problem, and improves the system's reliability and storage efficiency.
\subsubsection{Query Optimization}
SE faces a query efficiency bottleneck. Optimizing query algorithms and index structures can significantly enhance query performance. Guan et al. \cite{[1]} enhanced HE's encryption and decryption efficiency by optimizing modular exponentiation with a fast-exponentiation algorithm, reducing encryption and decryption times across different key lengths. Wang et al. \cite{[58]} introduced a scheme with lower time costs than traditional sequential-indexing and other robust technologies, particularly effective for numerous keywords. Florea et al. \cite{[96]} designed a customized coding method for network stream  data to boost computational efficiency. Xu et al. \cite{[3]} combined with overlapping clustering algorithm, introduced keyword weight sorting, optimized search path and sorting strategy, and improved search accuracy and efficiency. Hu et al. \cite{[26]} uses expressive boolean query, which does not increase the search time and the total number of outsourced documents linearly.
%, and supports the specification of access time period for DU to search through encrypted EHR, where once the time seal expires, the access right will be automatically revoked.
%Zhang et al. \cite{[77]} improved retrieval efficiency by introducing inverted index and least frequent keyword (s-term) technology, reducing search complexity from linear to sublinear. Wang et al. \cite{[79]} proposed a bidirectional index structure for efficient encrypted data search and sharing, allowing search and share operations using the same private key. Zhang et al. \cite{[112]} integrated ABE and SE to achieve efficient search and fine grained access control in multi-client environments.
\subsubsection{Lightweight SSE}
SE is continuously exploring performance optimization. With many lightweight SSE technologies like symmetric encryption and key aggregation, its practicality in complex scenarios is enhanced. Xu et al. \cite{[115]} combined dynamic searchable symmetric encryption with blockchain bit matrix storage, enabling efficient data retrieval and updates. Key aggregation technology reduces the number of keys, lowering computational and storage overheads and improving system performance. Oh et al. \cite{[81]} allows data owners to manage data and access control without a TTP, solving traditional encryption key management problems and boosting data sharing efficiency. Li et al. \cite{[76]} introduced public key authentication encryption, ciphertext update, and keyword search. Using proxy updated ciphertext, they generated constant size trapdoor, addressing the linear increase of trapdoor with the number of senders in PAEKS schemes. Yang et al. \cite{[78]} employed pairing free operations to simplify computations on lightweight devices and enhance efficiency. 
%Wang et al. \cite{[109]} designed a new trapdoor generation algorithm, allowing users to search all authorized files with one aggregate trapdoor, significantly improving search efficiency.

\subsection{Performance Improvement Based on PRE}
In recent years, researchers have follow two directions to enhance SE. One combines SE with IBE and BE to optimize key management and data transmission. The other introduces lightweight encryption, refines algorithm design, and uses trusted third parties, significantly lowering computational complexity and storage demands.
\subsubsection{PRE Combined with IBE}
IBE simplifies public key management by using user identities as public keys, though traditional IBE schemes have limitations in data sharing flexibility. PRE enables flexible access rights transfer for encrypted data, enhancing sharing flexibility and security. Integrating IBE and PRE aims to achieve more efficient, flexible, and secure data sharing. Pei et al. \cite{[29]} applied IBE to key distribution, binding user identity and public key via identity hashing. This simplified public key management and enhanced data sharing security. Yao et al. \cite{[73]} proposed a chosen ciphertext attack secure identity-based PRE scheme with single hop conditional delegation and multi-hop ciphertext evolution, supporting efficient cloud based data storage and transmission and optimizing cloud-service data processing. Zhou et al. \cite{[86]} introduced a CRF to counteract backdoor attacks and ensure data transmission and storage security in risky environments. Tan et al. \cite{[11]} controlled proxy-server re-encryption authority based on data attribute weights, constructing access policies that allowed only authorized users to decrypt ABE ciphertext. Liu et al. \cite{[46]} proposed the efficient and secure key issuing IBE protocol, ensuring user identity legitimacy and improving medical data security. Zhao et al. \cite{[322]} provided post-quantum security guarantees based on lattice cryptography, and employed IBE to achieve two-way identity matching-thereby safeguarding the sender’s identity privacy against access by other entities.
\subsubsection{PRE Combined with BE}
The broadcast proxy re-encryption (BPRE) technology based on broadcast encryption can effectively send data to multiple recipients in the form of broadcast, combined with fine-grained access control mechanism, which significantly improves the efficiency and security of data sharing in multi-user environment. Wang et al. \cite{[54]} presented a blockchain-based cross-domain data sharing scheme for edge assisted the Industrial Internet of Things, achieving flexible and secure cross domain sharing while safeguarding smart device privacy. Zhang et al. \cite{[201]} combined BPRE with identity-based broadcast encryption, using vehicle identity-based encryption to lower computational complexity and enabling multiple authorized users to decrypt the same ciphertext. Ge et al. \cite{[95]} proposed a revocable identity-based BPRE scheme. It supports flexible cloud environment data sharing and shields sensitive information like volunteer genetic data from unauthorized access.
\subsubsection{Lightweight PRE}
By designing a lightweight PRE algorithm, the speed of ciphertext conversion or decryption can be accelerated, thereby improving the efficiency of data sharing. Reddy et al. \cite{[31]} leveraged the efficient encryption and decryption performance and low resource consumption of AES to quickly process data in a lightweight PRE framework, which significantly enhanced the security and efficiency of data sharing. Wang et al. \cite{[32]} employed the elliptic curve encryption algorithm to design a lightweight PRE scheme suitable for resource-constrained IoT environments. Raghav \cite{[67]} utilized Bilinear Pairing to achieve efficient encryption and re-encryption operations, leading to improvements in computational cost, consensus time, latency, and throughput. Liu et al. \cite{[71]} proved the superiority of the proposed scheme in security, efficiency and functionality, especially in the case of repeated sharing of EMRs, with lower computational overhead and better overall performance. Yang et al. \cite{[36]} store the encrypted PHR in the IPFS cluster, and only the hash value from IPFS and the symmetric key encrypted with the patient's public key are kept in the smart contract account, which reduces the blockchain storage cost and ensures data consistency and integrity. Li et al. \cite{[41]} make the computational overhead in the encryption, re-encryption and decryption stages independent of the number of attributes through the carefully designed PRE key generation method, and keep a constant level.
\subsubsection{Based on Trusted Third Parties}
In the blockchain environment, PRE's encryption, decryption, and re-encryption operations, which are highly computationally complex, need to be verified via consensus mechanisms. This results in slow transaction processing and difficulties in meeting real-time requirements. To address the storage and performance challenges of PRE, researchers have proposed several solutions. Guo et al. \cite{[24]} introduced an innovative dual channel storage mechanism and composite key query method to boost data query efficiency, offering a viable solution for large scale data sharing. Wang et al. \cite{[32]} employed an on-chain and off-chain cooperative storage mechanism, storing metadata on the blockchain for effective data supervision and packaged data on cloud servers to conserve blockchain storage resources. Their data packaging mechanism allows edge servers to package data from smart devices, which reduces the number of interactions with the blockchain, thus reducing the computational and storage burdens of the blockchain. Wang et al. \cite{[34]} utilized B + tree indexes to enhance the query speed and stability of historical health data and optimize the user experience of data sharing.

\section{Function Improvement of Data Sharing}\label{s6}
ABE, SE, and PRE ensure data security and enhance the practicality and convenience of data sharing. This section systematically summarizes and analyzes existing research on how these technologies improve the functionality of data sharing, as shown in Table \ref{table4}. %\Lv{Reorder according to the order of classification and appearance}

\subsection{Function Improvement Based on ABE}
Achieving efficient, flexible, and controllable data sharing while ensuring data security and privacy has become a major challenge. With its unique advantages, ABE has become prominent in this field and has contributed to significant functional improvements.

\subsubsection{Access Control Management}
ABE can flexibly authorize or revoke a user's decryption rights, offering dynamic access control and fine-grained access management. It adapts to user group dynamics, boosting system flexibility and resource efficiency, and is widely used in fields like medical data sharing and cloud storage. Xu et al. \cite{[74]} introduced the concept of Revocable ABE, supporting dynamic user groups via timestamp management and enabling decryption authority revocation at the start of each time period. Cui et al. \cite{[68]} proposed an efficient revocation mechanism that updates only the proxy server's conversion key, not all user keys, reducing computational and communication costs. They also designed an anti-collusion attack mechanism to prevent revoked users from conspiring with the proxy server. Mei et al. \cite{[69]} leveraged puncturable encryption to let DO and DU independently delete specific data without third-party help. Ge et al. \cite{[85]} used ciphertext delegation so the cloud server could revoke ciphertext directly without re-encryption, lowering computation costs and enhancing system flexibility. Islam et al. \cite{[104]} enabled users to cancel or join multiple times via federated cloud architecture and enhanced tree-based group diffie-hellman trees without key redistribution, achieving efficient user management operations. Ge et al. \cite{[39]} designed an authorization and revocation mechanism to prevent malicious users from accessing and abusing data without authorization.

\renewcommand{\arraystretch}{0.1}
\begin{longtable}{
>{\raggedright\arraybackslash}p{0.8cm} 
>{\raggedright\arraybackslash}p{5cm} 
>{\raggedright\arraybackslash}p{1.5cm} 
>{\raggedright\arraybackslash}p{1.8cm}
>{\centering\arraybackslash}p{0.7cm} 
>{\centering\arraybackslash}p{0.9cm}
>{\centering\arraybackslash}p{0.8cm}}

\caption{Summary of Research on Improving the Function of Data Sharing by ABE, SE and PRE}
%\feng{Same} 
\label{table4} % 表格标签  
\\
\hline  
\footnotesize \textbf{Paper} & 
\footnotesize \textbf{Research Contents} & 
\footnotesize \textbf{Encryption Algorithm} & 
\footnotesize \textbf{Enhancement Technology}& 
\footnotesize \textbf{Secu\-rity} & 
\footnotesize \textbf{Perfor\-mance} & 
\footnotesize \textbf{Functi\-onality}  
\\

\hline
\endhead 
\addlinespace[2pt] %
\hline
\endfoot 
\hline
\addlinespace[2pt] %
\multicolumn{7}{l}{\footnotesize Note: The comparison of security strength,  performance and functionality is expressed as follows: \3 > \2 > \1.} 
%, indicating that variable \3 has the strongest influence on the dependent variable

\endlastfoot  
\addlinespace[2pt] %
\footnotesize\cite{[74]} & \footnotesize Access control management & \footnotesize ABE & \footnotesize CC & \1 & \1 & \3 \\
\footnotesize \cite{[68]} & \footnotesize Access control management & \footnotesize ABE & \footnotesize CC & \1 & \2 & \2 \\
\footnotesize\cite{[69]} & \footnotesize Access control management & \footnotesize ABE & \footnotesize CC & \1 & \1 & \3 \\
\footnotesize\cite{[85]} & \footnotesize Access control management & \footnotesize ABE & \footnotesize CC & \1 & \1 & \3 \\
\footnotesize\cite{[104]} & \footnotesize Access control management & \footnotesize ABE & \footnotesize CC & \1 & \1 & \3 \\
\footnotesize\cite{[39]} & \footnotesize Access control management & \footnotesize ABE, PRE & \footnotesize CC & \1 & \1 & \3 \\
\footnotesize\cite{[9]}&\footnotesize Traceability and audit &\footnotesize ABE, SE&\footnotesize BC&\2&\1&\2\\
\footnotesize\cite{[40]} & \footnotesize Traceability and audit & \footnotesize ABE, PRE & \footnotesize BC, IPFS & \2 & \2 & \3 \\
\footnotesize\cite{[42]}&\footnotesize Traceability and audit &\footnotesize ABE, PRE&\footnotesize BC, IPFS&\2&\1&\2\\
\footnotesize \cite{[17]} & \footnotesize Traceability and audit & \footnotesize ABE & \footnotesize CC, BC, Sig & \1 & \2 & \2 \\
\footnotesize\cite{[19]} & \footnotesize Traceability and audit & \footnotesize ABE & \footnotesize CC & \1 & \2 & \3 \\
\footnotesize\cite{[35]} & \footnotesize Traceability and audit & \footnotesize ABE & \footnotesize CC, Sig & \1 & \1 & \3 \\
\footnotesize \cite{[64]} & \footnotesize Cross-cipher system & \footnotesize ABE, PRE & \footnotesize CC, IBE & \1 & \2 & \2 \\
\footnotesize \cite{[11]} & \footnotesize Cross-cipher system & \footnotesize ABE, PRE & \footnotesize CC, IBE & \1 & \2 & \2 \\
\footnotesize\cite{[108]} & \footnotesize Cross-cipher system & \footnotesize ABE, PRE & \footnotesize IBE & \2 & \1 & \3 \\
\footnotesize\cite{[21]}&\footnotesize Cross-cipher system &\footnotesize ABE, PRE&\footnotesize CC&\2&\1&\2\\
\footnotesize\cite{[77]} & \footnotesize N-to-n scene, Static/dynamic search & \footnotesize SE & \footnotesize CC & \1 & \2 & \3 \\
\footnotesize\cite{[81]} & \footnotesize N-to-n scene, Multi-keyword search & \footnotesize SE & \footnotesize CC & \2 & \3 & \3 \\
\footnotesize\cite{[112]} & \footnotesize N-to-n scene, Multi-keyword search & \footnotesize SE & \footnotesize CC & \1 & \2 & \3 \\
\footnotesize\cite{[113]} & \footnotesize N-to-n scene & \footnotesize SE & \footnotesize CC & \1 & \1 & \3 \\
\footnotesize\cite{[115]} & \footnotesize N-to-n scene, Static/dynamic search & \footnotesize SE & \footnotesize BC & \2 & \3 & \3 \\
\footnotesize\cite{[114]} & \footnotesize N-to-n scene & \footnotesize SE & \footnotesize CC & \1 & \1 & \3 \\
\footnotesize\cite{[76]} & \footnotesize N-to-n scene, Multi-keyword search & \footnotesize SE, ABE& \footnotesize CC & \2 & \3 & \3 \\
%\footnotesize\cite{[91]} & \footnotesize N-to-n scene & \footnotesize SE & \footnotesize CC & \1 & \1 & \3 \\
\footnotesize\cite{[70]} & \footnotesize N-to-n scene & \footnotesize SE, PRE & \footnotesize BC & \2 & \1 & \2 \\
\footnotesize \cite{[96]} & \footnotesize N-to-n scene & \footnotesize SE & \footnotesize - & \1 & \2 & \2 \\
\footnotesize \cite{[321]} & \footnotesize N-to-n scene & \footnotesize SE & \footnotesize CC, PQC & \2 & \2 & \2 \\
\footnotesize \cite{[58]} & \footnotesize Multi-keyword search & \footnotesize SE & \footnotesize CC & \1 & \2 & \3 \\
\footnotesize \cite{[316]} & \footnotesize Multi-keyword search & \footnotesize SE & \footnotesize CC, PQC & \2 & \2 & \2 \\
\footnotesize\cite{[79]} & \footnotesize Static/dynamic search & \footnotesize SE & \footnotesize - & \2 & \2 & \2 \\
\footnotesize\cite{[98]} & \footnotesize Static/dynamic search & \footnotesize SE, ABE & \footnotesize - & \1 & \2 & \3 \\
\footnotesize \cite{[24]} & \footnotesize IPFS distributed storage & \footnotesize PRE & \footnotesize IPFS, BC, ZKP & \2 & \2 & \2 \\
\footnotesize\cite{[53]} & \footnotesize IPFS distributed storage & \footnotesize PRE & \footnotesize BC, IPFS & \2 & \1 & \3 \\
\footnotesize\cite{[62]}&\footnotesize IPFS distributed storage &\footnotesize PRE&\footnotesize BC&\2&\1&\2\\
\footnotesize\cite{[51]} & \footnotesize IPFS distributed storage & \footnotesize PRE & \footnotesize BC, IPFS & \2 & \1 & \3 \\
\footnotesize\cite{[83]} & \footnotesize IPFS distributed storage & \footnotesize PRE & \footnotesize BC & \1 & \1 & \3 \\
\footnotesize\cite{[57]} & \footnotesize Multi-hop PRE & \footnotesize PRE & \footnotesize CC & \1 & \1 & \3 \\
\footnotesize\cite{[63]} & \footnotesize Multi-hop PRE, One-to-many data sharing & \footnotesize PRE & \footnotesize IBE & \1 & \1 & \3 \\
\footnotesize \cite{[73]} & \footnotesize Multi-hop PRE & \footnotesize PRE & \footnotesize CC, IBE & \1 & \2 & \2 \\
\footnotesize\cite{[48]} & \footnotesize One-to-many data sharing& \footnotesize PRE, ABE & \footnotesize - & \1 & \1 & \3 \\
\footnotesize\cite{[28]} & \footnotesize One-to-many data sharing& \footnotesize PRE & \footnotesize IBE, BE & \1 & \1 & \3 \\
\footnotesize\cite{[89]} & \footnotesize One-to-many data sharing& \footnotesize PRE & \footnotesize CC, BC & \2 & \1 & \3 \\
\footnotesize\cite{[54]} & \footnotesize One-to-many data sharing& \footnotesize PRE & \footnotesize BC, BE & \1 & \2 & \2 \\

%\footnotesize\cite{[23]} & \footnotesize Access control management & \footnotesize ABE & \footnotesize BC & \2 & \1 & \3 \\
%\footnotesize\cite{[45]} & \footnotesize Access control management & \footnotesize ABE & \footnotesize BC, CC & \2 & \0 & \3 \\

%\footnotesize\cite{[100]} & \footnotesize Traceability and audit & \footnotesize ABE & \footnotesize BC & \2 & \1 & \3 \\
%\footnotesize\cite{[110]} & \footnotesize Traceability and audit & \footnotesize ABE & \footnotesize BC, CC & \2 & \1 & \2 \\

\end{longtable}
\renewcommand{\arraystretch}{1.0}

\subsubsection{Traceability and Auditing}
In the intricate landscape of data sharing, traceability and auditing are vital for ensuring data safety throughout its lifecycle. Blockchain technology, with its tamper proof and distributed ledger properties, meticulously records all data sharing activities. This provides a robust foundation for data traceability and auditing. When combined with ABE, it enables fine-grained access control while logging all access requests and authorization details. This integration ensures the security, compliance, and reliability of data sharing. Extensive studies, including \cite{[9],[40],[42]}, have demonstrated the efficacy of this technological combination. These works show how smart contracts and blockchain technology can be deeply integrated to achieve data sharing traceability and auditability while protecting data privacy.\par
Unlike blockchain's approach to data traceability and auditing, many existing studies rely on cryptographic methods such as homomorphic signatures and certificateless signatures to verify data integrity and authenticate identities. Yao et al. \cite{[17]} employed homomorphic signatures to sign each data block during data integrity audits. This allowed verification of the validity of certificates returned by the cloud server. A third-party auditor could then conduct efficient audits to ensure data integrity, avoiding complex zero-knowledge interactive proofs and aggregate signatures. Guan et al. \cite{[19]} used certificateless authentication technology to audit the integrity of shared data files. This method sidestepped the certificate issues in PKI and the key escrow attack risks in IBE. Zhao et al. \cite{[35]} proposed a scheme combining threshold authorization with traceability. Data owners could delegate multiple authorized users to complete authorization operations in a threshold-based manner, enabling the tracing of authorized users during the authorization process. This ensured the transparency and controllability of the authorization process.\par

\subsubsection{Cross-Cipher System}
In cross-domain data sharing, PRE technology stands out for its efficient ciphertext conversion capabilities. When combined with ABE, it enables a one-to-many data sharing model, significantly enhancing data sharing efficiency. However, differences in cryptographic systems across trust domains pose challenges for ABE interoperability and cross-domain data sharing. Wang et al. \cite{[64]} proposed the first unbounded cross-domain PRE scheme to transform IBE ciphertext into ABE ciphertext. Tan et al. \cite{[11]} used PRE technology to convert IBE ciphertext into ABE ciphertext based on data attributes and their weights, achieving precise control over proxy server re-encryption authority. Deng et al. \cite{[108]} introduced the Hybrid Attribute-Based Proxy Re-encryption (HAPRE) encryption paradigm, solving the problem of accessing ABE-encrypted data on resource-limited devices. They constructed two HAPRE schemes, proving their semantic security and anti-collusion capabilities. Sun et al. \cite{[21]} proposed a verifiable and hybrid attribute-based PRE, which transforms ABE ciphertext into PKI ciphertext using key blinding technology to protect ABE users' private keys and prevent collusion attacks between cloud servers and PKI users.

\subsection{Function Improvement Based on SE}
By introducing N-to-N scenarios, SE allows multiple users to efficiently retrieve multiple datasets. Its multi-keyword search enhances query flexibility and precision. Moreover, optimizations in static and dynamic search functions enable SE to better adapt to data dynamics, thereby improving the system's overall performance.
\subsubsection{N-to-N Scenario}
SE faces functional limitations in complex environments with multiple users and datasets. N-to-N scenario optimization enables SE to support efficient multi-user retrieval of multiple datasets, significantly enhancing system flexibility and usability. References \cite{[77],[81],[112],[113],[115]} describe a retrieval scheme supporting many-to-many communication. It allows multiple senders to encrypt data for simultaneous retrieval by multiple receivers. Li et al. \cite{[114]} proposed a secure cloud data sharing protocol with hierarchical keyword search. This enables multiple DO to share data with multiple users, with receivers accessing different data levels based on their authority. Li et al. \cite{[76]} present a medical data sharing scheme for interactions between multiple DO and DU. It supports multi-user data sharing and access control. Zhou et al. \cite{[70]} explored vehicle social networks and secure data sharing using blockchain and encryption. In one-to-many scenarios, DO need to share data safely with multiple users while ensuring data privacy and integrity. Florea et al. \cite{[96]} examined encrypted-search schemes based on symmetric and asymmetric keys. This is vital for one-to-one scenarios where DO and query clients aim to enhance search efficiency without compromising data privacy. Xu et al. \cite{[321]} constructed an efficient PAEKS scheme with quantum resistance, multi-user support and forward security.
\subsubsection{Multi-Keyword Search}
SE enables accurate and efficient ciphertext environment retrieval through multi-keyword search, enhancing the integration of data security and search functionality. Wang et al. \cite{[58]} detailed a multi-keyword search scheme based on zero-sum confusion Bloom filters and oblivious transfer, ensuring result unlinkability. Li et al. \cite{[76]} introduced a non-colluding agent to update ciphertexts, achieving constant size trapdoor generation and solving the traditional issue of trapdoor size increasing linearly with sender numbers. Oh et al. \cite{[81]} supported multi-keyword search with a bidirectional structured index. Users can directly match encrypted search indexes with search tokens and use index chains to locate relevant data, improving multi-keyword search efficiency. Zhang et al. \cite{[112]} proposed a multi-client Boolean keyword search scheme. Its search complexity depends on the number of documents with the least frequent keywords, not the total documents stored, achieving sublinear complexity and boosting search efficiency. Yao et al. \cite{[316]} proposed an ideal lattice-based SE scheme that not only achieves efficient multi-keyword retrieval but also guarantees keyword privacy in both the ciphertext and the search token.
\subsubsection{Static/Dynamic Search}
SE can satisfy retrieval needs in both static and dynamic data environments through static and dynamic search functions, enhancing retrieval efficiency and flexibility. Static SE, being simpler and more efficient, suits straightforward scenarios like data archiving. Zhang et al. \cite{[77]} proposed the DSB-SE scheme, which redesigns traditional Boolean SSE with a new transformation key module. This converts data-reader query formats to data writer query formats, achieving search speeds tens of thousands of times faster than conventional schemes. However, static SE's inability to modify uploaded data spurred the development of dynamic SE, which allows post upload data modifications but introduces security challenges like forward and backward privacy protection. References \cite{[79]} and \cite{[98]} advanced a SE scheme enabling efficient data search and sharing while safeguarding forward and backward privacy. Xu et al.  \cite{[115]} introduced a blockchain-based dynamic SSE scheme. It ensures high retrieval and update efficiency, breaks down data barriers, and enhances the reliability and security of data sharing.
%Li et al. \cite{[76]} constructed an inverted index to quickly locate ciphertexts with specific keywords post update, without a linear search of the entire ciphertext space. 
\subsection{Function Improvement Based on PRE}
To address traditional storage bottlenecks, the introduction of IPFS can realize efficient storage and secure sharing of encrypted data. Multi-hop PRE technology breaks through the single-hop limit and enhances the flexibility and fault tolerance of the system. PRE combines ABE and other technologies to expand its application in one-to-many data sharing scenarios. This section combs the relevant research progress and analyzes the innovation mechanism.
\subsubsection{Distributed Storage Based on IPFS}
IPFS provides a reliable foundation for the storage and management of encrypted data by virtue of its distributed storage architecture, efficient storage space utilization and data persistence. It cooperates with PRE, which significantly improves data security and access flexibility by storing encrypted data, realizing fine-grained access control and efficient data sharing and distribution. Guo et al. \cite{[24]} stored the encrypted medical data in IPFS in the dual-channel storage mechanism, which effectively isolated the medical data from the patient identity information and reduced the risk of privacy disclosure. Yeh et al. \cite{[53]} combined the mechanism of Editable Blockchain and revocable IPFS for the first time, which overcame the problem that data in traditional blockchain could not be deleted and realized the function of completely deleting PHR. Mittal et al. \cite{[62]} used IPFS technology to securely store medical data offline, providing high throughput and concurrent access. References \cite{[51]} and \cite{[83]} combine blockchain with IPFS, and with the distributed storage advantages of IPFS, the problems in data storage and sharing are solved, and the overall performance and security of the system are significantly improved.
\subsubsection{PRE Based on Multi-Hop}
In the PRE domain, multi-hop technology stands out for its ability to incrementally transform ciphertext via multiple intermediate proxy nodes. This approach not only enhances the flexibility and security of the system but also effectively reduces the burden and risk associated with a single proxy node. Luo et al. \cite{[57]} introduced a multi-hop re-encryption scheme. Their scheme allows for numerous ciphertext transformations, catering to the complexities of cloud environments. It delivers clear advantages in computational and communication efficiency. Yang et al. \cite{[63]} developed a scheme that combines multi-hop capabilities with one-way properties and resistance to collusion attacks. This makes it well-suited for multi-user scenarios while bolstering data security and privacy. Yao et al. \cite{[73]} put forward the first CCA-secure identity-based PRE scheme. This innovative scheme supports both single-hop conditional delegation and multi-hop ciphertext evolution, thus filling a critical gap in the field.
\subsubsection{One-to-Many Data Sharing}
ABE, IBE, and BE are advanced encryption technologies with great potential for efficient, flexible one-to-many data sharing. They offer diverse solutions through different mechanisms, meeting fine-grained access control requirements and enhancing data sharing efficiency and scalability. Zhao et al. \cite{[48]} proposed a new attribute-based threshold PRE scheme. It addresses the single point of failure in traditional schemes, boosting system robustness and security, and provides a safer, more flexible solution for fine-grained access control and data delegation. References \cite{[28],[63],[89]} integrated PRE and IBE, using user identity information as public keys. This reduces public key management complexity and supports one-to-many data sharing. Wang et al. \cite{[54]} designed a data sharing protocol combining BE and PRE. It leverages BE's efficient one-to-many transmission characteristics for batch data distribution to multiple authorized users, significantly improving data sharing efficiency and scalability.
%Reddy et al. \cite{[31]} combined PRE and ABE with access control policies to enable fine grained data access, allowing only users with specific attributes to decrypt data. 
\section{Application of Data Sharing}\label{s7} %\Lv{The whole section is completely rewritten, and the application classification of 115 articles cited in this paper is counted, and Figure 4 is made. According to the application classification, the introduction is written and the references are added.} \par
\begin{figure*}[!t]
    \centering
    \includegraphics[width=13cm]{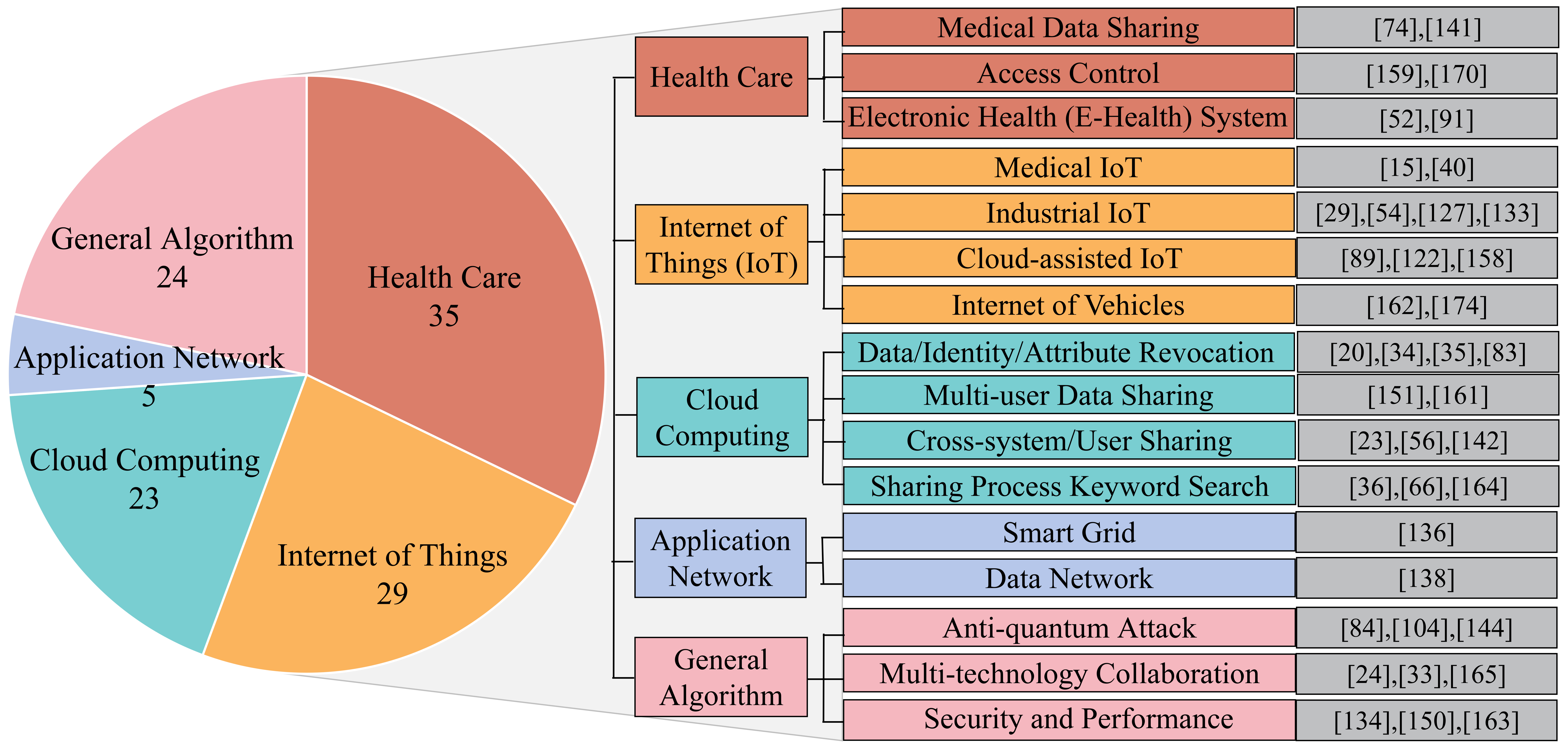}
    \caption{Application field distribution of cited papers and recommendation of representative works}
    \label{p5}
\end{figure*}

%ABE, SE, and PRE not only guarantee data security but also boost the efficiency and flexibility of data sharing, strongly driving the digital transformation across various industries. Fig.\ref{p5} (a) illustrates the application field distribution of the 116 papers analyzed in this study, of which 25 papers are general algorithms. Fig.\ref{p5} (b) presents some cutting-edge research in various application fields.
Privacy-enhancing encryption technologies not only secures data security but also boosts data sharing efficiency and flexibility, driving digital transformation across industries. Fig.\ref{p5} presents the application field distribution of the 116 analyzed papers and recommends advanced papers in related fields.

%ABE, SE, and PRE are not only focuses of academic research but also play crucial roles in practical applications. They ensure data security, enhance the efficiency and flexibility of data sharing, and strongly support the digital transformation of various industries. The following is a detailed introduction to the cutting-edge applications of ABE, SE, and PRE in different fields. Fig.\ref{p5} \feng{need citations. Better add a table}

\begin{itemize}
\item[1)] Health Care: The secure sharing of patient records, images, and other medical data across institutions is essential, and patient privacy must be rigorously protected. When sharing medical data, PRE allows for the secure transfer of patient data to third parties for analysis without exposing sensitive information in plaintext \cite{[71],[72]}. Hospitals and research institutions can dynamically grant access based on the attributes of doctors and other entities \cite{[53],[56]}. Within electronic health systems, SE enables the quick retrieval of keywords from encrypted medical records, further supporting efficient and secure data management \cite{[22],[26]}.
\item[2)] Internet of Things: IoT devices generate massive data requiring sharing across platforms and users. Given the limited resources of IoT devices, designing a lightweight encryption scheme is crucial for large-scale data sharing, ensuring data security and low-power device operation. In medical IoT, SE protects data privacy and boosts search efficiency \cite{[1]}, and ABE-based access control ensures safe sharing of medical wearable data \cite{[5]}. In industrial IoT, designing multi-chain cloud-chain architectures or optimizing search algorithms can improve data search security and enable safe, flexible data sharing \cite{[12],[18],[32],[54]}. For cloud-assisted IoT, combining ABE, SE, and PRE protects device data privacy and supports cross-platform secure sharing and rapid response \cite{[11],[17],[69]}. In internet of vehicles, schemes supporting lightweight search or multi-user data sharing are suitable for scenarios like vehicle formation driving and traffic information sharing \cite{[70],[201]}.
\item[3)] Cloud Computing: Centralized data storage and on-demand sharing are key features that also pose significant data security and privacy challenges. ABE enables flexible revocation of data, identities, and attributes \cite{[39],[57],[68],[85]}. Once a user's rights or attributes change, promptly updating the encryption strategy can stop data leakage to unauthorized users and secure data. ABE can streamline multi-user data sharing in enterprise team-work scenarios, simplifying the entire process \cite{[51],[63]}. PRE makes cross-system and cross-user sharing possible \cite{[104],[74],[84]}. It breaks down encryption barriers between different systems and user groups, thereby promoting secure and efficient data interaction. SE facilitates keyword search during data sharing. Users can directly search encrypted data in the cloud, quickly find needed information, boost retrieval efficiency, and ensure data privacy \cite{[112],[113],[114]}.
\item[4)] Application Network: In the realm of smart grids, PRE serves as a powerful tool for safeguarding privacy and enhancing robustness. It ensures that data aggregation remains accurate even in the face of equipment failures and allows for universal verification of the aggregated results \cite{[47]}. In the context of named data networking, the integration of PRE with symmetric mechanisms offers a comprehensive solution. This combination not only protects the privacy of content but also prevents the leakage of ciphertext copies and sensitive information contained in content names \cite{[55]}. 
\item[5)] General Algorithm: The general algorithm is not designed for a specific field, but focuses on the bottom optimization of algorithm technology. At present, encryption technology is generally threatened by quantum attacks, and it is an effective technical path to study privacy-enhancing encryption technology against quantum attacks \cite{[59],[321],[323]}. Multi-technology cooperation can break through the inherent bottleneck of single encryption technology in function, performance and security, and meet the multi-dimensional requirements of flexibility and expansibility \cite{[7],[44],[101]}. The general algorithm with balance between security and performance can effectively reduce the calculation, communication and storage expenses of the algorithm under the premise of ensuring the confidentiality and integrity of data by means of lightweight cryptographic operations, and realize the cooperative consideration of the security protection ability and performance of the scheme \cite{[77],[78],[79]}.
\end{itemize}
\par

\section{Discussion, Challenge and Future Direction}\label{s8}
This section discusses the problems in the existing data sharing schemes and summarizes the challenges in solving these problems. Finally, this work points out the future research direction for data sharing based on privacy-enhancing encryption technologies.
%\feng{which are SOTAs for each problem? Lack necessary insights for authors. }
%\subsection{Discussion and Challenge} 
%\Lv{Add citations} 
%\feng{Cite after the term: Algorithm security []}
%Security Improvement Problem
\begin{itemize}
\item[1)] The optimization of data sharing scheme based on quantum security:\par 
%The lack of anti-quantum capability is the core underlying security flaw in current data sharing schemes \cite{[101],[17],[68],[69],[71],[72],[57]} based on SE, PRE, and ABE, rooted in their reliance on classical mathematical problems such as discrete logarithm and bilinear pairing. These problems can be efficiently cracked by Shor and Grover algorithms once quantum computing matures \cite{[318],[319],[320]}. In terms of technical characteristics, the trapdoor generation and index matching of SE \cite{[321]}, re-encryption key derivation and ciphertext conversion of PRE \cite{[322]}, and attribute trapdoor construction and decryption verification of ABE all depend on the above mathematical primitives to achieve ``a balance between security and functionality'' \cite{[323]}, which further exacerbates quantum security risks. \par
% To address this issue, future efforts will focus on lattice-based cryptography optimization: the SE scheme \cite{[55]} can construct a trapdoor and index mechanism based on RLWE, leveraging lattice-based quantum resistance to ensure security \cite{[316]}. The PRE scheme \cite{[12],[7]} adopts ideal lattices or NTRU lattices to design anti-quantum re-encryption keys, guaranteeing the confidentiality of keys and ciphertexts. \cite{[317]}The ABE scheme \cite{[59]} needs to break through lattice-based attribute trapdoor construction technology, utilizing lattice-based homomorphic properties to match policies and attributes, while resisting Grover algorithm attacks to ensure verification security.

The lack of anti-quantum capability is the core underlying security flaw in current data sharing schemes based on SE, PRE, and ABE, rooted in their reliance on classical mathematical problems such as discrete logarithm and bilinear pairing \cite{[101],[17],[68],[69],[71],[72],[57]}. These problems can be efficiently cracked by Shor and Grover algorithms once quantum computing matures \cite{[318],[319],[320]}. In terms of technical characteristics, the trapdoor generation and index matching of SE \cite{[321]}, re-encryption key derivation and ciphertext conversion of PRE \cite{[322]}, and attribute trapdoor construction and decryption verification of ABE all depend on the above mathematical primitives to achieve ``a balance between security and functionality'' \cite{[323]}, which further exacerbates quantum security risks. \par
To address this issue, future efforts will focus on lattice-based cryptography optimization: The SE scheme can construct a trapdoor and index mechanism based on RLWE, leveraging lattice-based quantum resistance to ensure security \cite{[316]}. The PRE scheme adopts ideal lattices to design anti-quantum re-encryption keys, guaranteeing the confidentiality of keys and ciphertexts \cite{[322]}. The ABE scheme needs to break through lattice-based attribute trapdoor construction technology, utilizing lattice-based homomorphic properties to match policies and attributes, while resisting Grover algorithm attacks to ensure verification security \cite{[317],[326]}.

%\begin{itemize}
%    \item Algorithm security \cite{[101],[17],[68],[69],[71],[72],[57]}: If the defense mechanism of encryption algorithm is flawed, it will easily lead to data leakage, privacy invasion and malicious tampering or attack on the system, which will lead to information distortion, decision-making mistakes and even a trust crisis. Therefore, it is undoubtedly a challenging research topic to explore how the existing encryption algorithms can ensure the security of data sharing in multiple and complex scenarios.\par 
%- Key management security: There are hidden security risks in the key generation, distribution, storage and update. Once the key is leaked, the data security will face severe threats. Therefore, it has become an important research challenge to explore how to build a more secure and efficient key management mechanism to ensure the security of keys throughout their life cycle.\par 
%    \item Data security \cite{[55],[12],[7],[59]}: Encryption operation may reveal side channel information such as execution time and memory access mode, which can be used by attackers to carry out side channel attacks and infer keys or sensitive data, posing a threat to data security. In the design of encryption algorithm and the implementation of the system, it is a challenging task to take targeted measures to ensure its security.\par 
%\end{itemize}

\item[2)]
Security and performance optimization of data sharing in low-resource scenarios: \par 
In smart homes, smart healthcare, smart transportation, and other scenarios with low computing power and limited storage, the balance between the efficiency and security of data sharing is particularly prominent. In the face of large-scale data processing scenarios, the problem of excessive computational overhead arises frequently \cite{[18],[77],[79],[47]}. How to optimize algorithm design and reduce the time complexity of encryption, decryption, and search operations has become a core research issue that urgently needs to be addressed \cite{[56]}. Meanwhile, the storage cost of encrypted data is much higher than that of plaintext data, which imposes significant pressure on specific scenarios with limited storage resources \cite{[32],[54]}. In inter-agency data sharing scenarios, efficient data transmission and sharing are highly dependent on reliable communication mechanisms \cite{[76],[78],[70],[55]}. Therefore, optimizing communication protocols, reducing transmission latency, and improving the real-time performance and response speed of sharing are also key directions in this field.\par 
To address the above dilemmas, it is necessary to construct an integrated solution from the two dimensions of security and performance optimization, combined with cutting-edge technologies. In terms of security enhancement: first, adopt a multi-algorithm fusion strategy to integrate the high efficiency of symmetric encryption and the high security advantages of asymmetric encryption, thereby designing a composite encryption scheme that balances efficiency and security \cite{[55],[58]}. Second, leverage blockchain and smart contract technologies to achieve decentralized control and automatic execution of data sharing, further enhancing system security and operational transparency \cite{[53]}. Third, introduce advanced technologies such as differential privacy and data anonymization to upgrade the privacy protection mechanism, ensuring that user privacy information is not leaked throughout the entire process of data sharing and retrieval \cite{[44]}. In terms of performance optimization: First, develop lightweight encryption algorithms suitable for resource-constrained environments such as IoT devices and mobile terminals to meet the application requirements of scenarios with low computing power and limited storage \cite{[97]}. Second, accurately optimize key lengths on the premise of meeting core security requirements, avoiding additional computational overhead and storage burdens caused by excessively long keys \cite{[41]}. Third, establish an efficient key distribution and dynamic update mechanism, which significantly reduces the complexity of key management and provides support for the efficient promotion of data sharing \cite{[38]}.

\item[3)] Building a trusted data protection system for the entire data lifecycle via multi-technology integration: \par
%Combining privacy-enhancing encryption technology with enhancement technology can establish a credible protection system covering the entire lifecycle of data collection, transmission, storage, and usage. The core advantage of this technological integration is that it not only retains the core capabilities of privacy-enhancing encryption technology in fine-grained control, ciphertext retrieval, and authority transfer but also addresses the shortcomings of traditional encryption schemes in data availability, multi-agent collaboration, and the depth of privacy protection by integrating cutting-edge privacy protection technologies. It can be widely applied to data sharing scenarios in medical, government, financial, industrial, and other fields, providing robust technical support for the safe and orderly flow of data elements.\par 
%This serves as the core approach to achieving the principle of ``balancing security and efficiency, privacy and value'' in data sharing. 
Existing research on constructing a trusted data protection system covering the entire data lifecycle through multi-technology integration still faces numerous unsolved bottlenecks, specifically as follows: First, the depth of multi-technology integration is inadequate, making it challenging to build an integrated full-lifecycle trusted data protection system. Second, existing schemes struggle to retain the core capabilities of privacy-enhancing encryption while balancing data availability, multi-agent collaboration efficiency, and the depth of privacy protection, failing to effectively break through the core bottlenecks of traditional encryption schemes. Third, technical solutions have limitations in scenario adaptability. For data sharing requirements across diverse fields such as healthcare, government, and enterprise, there is a lack of universal and accessible solutions, hindering the safe and orderly flow of data.\par 
The future development direction can be divided into three core levels: First, at the data acquisition and preprocessing level, differential privacy technology is integrated to perturb raw data \cite{[310]}. ABE technology is also used to encrypt the perturbed data. Only authorized users can obtain the original perturbation rules and encrypted data, thereby achieving a balance between data privacy and availability. This is suitable for large-scale sharing scenarios \cite{[311]}. Second, at the data computation and analysis level, homomorphic encryption and federated learning technologies are integrated. In multi-agent collaborative data computing scenarios, there is no need for all agents to share raw data: homomorphic encryption technology enables authorized parties to directly compute on encrypted data, with the calculation results remaining accurate after decryption \cite{[312],[324]}. Relying on PRE technology, the secure circulation of encrypted computing authority is realized, supporting multi-agent relays to complete complex computing tasks \cite{[313]}. Combined with the federated learning framework, all participants retain raw data locally and only encrypt model parameters for sharing via SE technology, achieving ``data available but invisible'' and effectively solving the problem of privacy leakage in cross-enterprise and cross-domain collaborative analysis \cite{[314]}. Third, at the data storage and access level, blockchain and privacy computing technologies are integrated \cite{[314]}. A distributed data sharing platform is built using blockchain, where key information such as ABE's permission policies, SE's retrieval indexes, and PRE's re-encryption records are uploaded to achieve full traceability and auditability of the data sharing process \cite{[315]}. Privacy computing nodes conduct real-time verification of visitors' identities and permissions, and smart contracts automatically enforce access control rules, further enhancing the credibility of data sharing \cite{[312],[325]}.

\item[4)] 
Cross-institutional data collaboration boosts the development of emerging industries: \par
Against the AI backdrop, its integration with medical, financial and transportation industries is significant. However, in practice, there are problems such as difficulties in establishing trust between institutions, and system heterogeneity, including inconsistencies in cryptographic systems, data formats, interface standards, and communication protocols, further raises the technical threshold for data sharing.\par
In the future, it is necessary to build a trust mechanism and strengthen inter-institutional mutual trust to fully release the data value of ``AI + Industry'', which can be achieved through the following paths: First, formulate reasonable incentive measures to stimulate the enthusiasm of various institutions to participate in data sharing \cite{[54]}. Second, develop heterogeneous system integration schemes and design solutions that enable secure cross-system communication based on the differences between different systems \cite{[64]}. Third, utilize middleware technology to develop and deploy cross-system middleware, complete data format conversion and communication protocol adaptation, and solve system compatibility issues \cite{[7]}.\par
\end{itemize}
\par

%\subsection{Future Direction} 
%\Lv{Add citations} 
%\feng{need citations. Initially prove your proposed directions and show the oppotunities}

\section{Conclusion}\label{s9}
This work systematically reviews and organizes existing data sharing schemes based on privacy-enhancing encryption technologies. Firstly, a refined analytical framework is constructed, the complete process of data sharing is outlined, and various potential security threats are comprehensively analyzed. Secondly, focusing on the three core dimensions in the field of data sharing: security threats, performance bottlenecks, and functional adaptability. 12 enhancement technologies are summarized and integrated with privacy-enhancing encryption technologies. The research progress and practical outcomes of data sharing schemes based on this technology combination are thoroughly discussed with respect to the above three dimensions. Finally, we summarize and refine the existing advanced solutions capable of delivering privacy protection, intelligent enhancement, and cross-domain data collaboration across diverse scenarios. Combined with forward-looking research directions, this work offers valuable theoretical foundations and practical insights for the subsequent development and application of data sharing technologies.
%This paper presents a thorough review of existing data-sharing mechanisms that are based on privacy-enhanced encryption technology. We establish a detailed framework outlining the data sharing process and examine potential security threats. The research centers on three critical aspects of data sharing: security, performance bottlenecks, and functional adaptability. We explore in depth the significant role that privacy-enhanced encryption technology plays in these mechanisms. In addition, we investigate 11 commonly used enhancement technologies. By addressing the security threats and performance limitations in data sharing and incorporating forward-looking research directions, we aim to offer valuable reference and inspiration for the development and application of data-sharing technologies.

%\section{Acknowledgments}
%This work was supported by National Key Research and Development Program of China (2023YFB3107100).

%\bibliographystyle{ACM-Reference-Format}
%\bibliography{sample-base}

\end{document}